%% file: main.tex
\documentclass[twocolumn, astrosymb]{aastex701}
\pdfoutput=1 

\usepackage{amsmath,amstext}
\usepackage{graphicx}

\input{defs.tex}

\received{August 10, 2026}

\submitjournal{\apj}

\shorttitle{Carbon Giants Revealed by Gaia DR3}
\shortauthors{Roulston et al.}

\turnoffediting

\begin{document}

\title{The Carbon Giant Population Revealed by Gaia DR3}

\author[0000-0002-9453-7735]{Benjamin R. Roulston}
\affiliation{Department of Physics, Clarkson University, 8 Clarkson Ave, Potsdam, NY 13699, USA}
\affiliation{Institute for STEM Education, Clarkson University, 8 Clarkson Ave, Potsdam, NY 13699, USA}
\email{broulsto@clarkson.edu}
\correspondingauthor{Benjamin R. Roulston}
\email{broulsto@clarkson.edu}

\author[0000-0002-8179-9445]{Paul J. Green}
\affiliation{Center for Astrophysics $\vert$ Harvard \& Smithsonian, 60 Garden Street, Cambridge, MA 02138, USA}
\email{pgreen@cfa.harvard.edu}





\begin{abstract}
Classical carbon stars (atmospheric C/O$>1$) are asymptotic giant branch (AGB) stars that become carbon-rich through third dredge-up during the thermally pulsing AGB phase. However, carbon stars also occur among red giants and main-sequence stars, where the carbon is thought to originate from binary mass transfer from a former AGB companion. Using the all-sky catalog of $G<16.5$,mag \GDR3\ carbon stars from \citet{Roulston2025}, we investigate a uniform, high-confidence sample of carbon giants in the Milky Way, its globular clusters, and nearby Local Group galaxies. We identify 6937 carbon giants in the Large Magellanic Cloud, 2148 in the Small Magellanic Cloud, and 219 associated with the Sagittarius dwarf spheroidal, nearly quadrupling the previously known sample in the latter. We identify several candidate carbon symbiotic stars through their X-ray counterparts. The mean absolute C-AGB magnitude, $M^0_J=-6.30$ in the LMC and $-6.15$ in the SMC, agrees closely with the photometrically derived values of \citet{Ripoche2020}, reinforcing the utility of carbon AGB stars as distance indicators. From 4276 Galactic carbon giants with $M_G<0$ and $|b|>5^\circ$, we measure a local disk space density of $n_0=11\pm0.5\times10^{-8},\mathrm{pc}^{-3}$ and a scale height of $H_z=257\pm4$ pc. Lower-luminosity carbon giants exhibit significantly larger \GDR3\ RUWE values than C-AGB stars, providing strong evidence that most are extrinsic post-mass-transfer systems, likely evolved dwarf carbon stars. C-AGB stars are more tightly confined to the Galactic disk and are rare inside $R\lesssim7$ kpc, consistent with Galactic age and metallicity gradients, whereas the lower-luminosity carbon giants have a larger scale height and require an additional halo component.
\end{abstract}



\section{Introduction \label{sec:intro}}

Carbon-rich stars form a diverse and complex population, encompassing objects across a broad range of evolutionary states. 
Strong atmospheric carbon (C$>$O) is typically found in asymptotic giant branch (AGB) stars that have experienced the third dredge-up during the thermally pulsing AGB (TP-AGB) phase, when deep convective mixing transports helium, carbon, and $s$-process elements from the stellar interior to the surface \citep{Iben1974, Iben1983}. These TP-AGB carbon (C) stars are luminous and can be observed out to great distances, in the Milky Way halo, the Large and Small Magellanic Clouds, and nearby galaxies.

However, most carbon stars are not in the TP-AGB phase. The carbon star family includes a range of classes traditionally classified as C-N, C-J, and C-R types, as well as notable subgroups like Barium (Ba), C-H, and carbon-enhanced metal-poor (CEMP) stars \citep{Wallerstein1998}. C-N stars, which often include AGB stars and luminous giants, display the deepest carbon bands and reddest colors. In contrast, C-H and C-R stars are typically bluer, with weaker carbon bands and are often classified as subgiants. C-H and C-R stars can be difficult to distinguish from one another, and many stars historically labeled as C-R are likely misclassified C-N or C-H stars. The true evolutionary origins of early-type C-R stars are still under debate \citep{Izzard2007, Zamora2009}.

Barium stars also exhibit relatively weak carbon bands, similar to those in early C-R stars, but can be distinguished spectroscopically by prominent Ba\,II absorption lines at 4007\AA\ and 4554\AA. A common feature among C-H, Ba, and $s$-process-rich CEMP-$s$ stars is their high binary fraction \citep{McClure1990, Jorissen1998, Lucatello2005, Jorissen2016}, which suggests an extrinsic origin for their C enrichment - previous mass transfer from an AGB C star.

Not all C stars are giants. Carbon-rich main sequence stars are created via mass transfer in binary systems. During the short-lived TP-AGB phase \citep[lasting only a few million years;][]{Kalirai2014}, some AGB C stars can transfer C-enriched material to their main sequence companions. After this mass transfer, the AGB star loses its envelope and becomes a white dwarf, which becomes dimmer than its main sequence companion on the order of $\sim 10$\,Myr. The result is a main sequence star with a carbon-enriched atmosphere - what we recognize as a dwarf carbon (dC) star. Depending on the mass and metallicity of the companion, the amount and composition of accreted material, and the depth of envelope mixing, C$_2$ and CN molecular bands may become prominent even in low- or medium-resolution spectra \citep{Dahn1977, Green1991, Christlieb2001, Li2018}.  As dC stars reflect the integrated history of mass transfer, they have a larger space density than all the C giants combined \citep{Roulston2025}.  

Supporting this post-mass-transfer scenario, some DA/dC double-lined spectroscopic binaries still show the hot white dwarf in their spectra \citep{Heber1993, Liebert1994, Si2014}. Observationally, the binary fraction among dC stars is remarkably high—around 95\% \citep{Roulston2019}. As these carbon-enriched dwarfs evolve off the main sequence, they may become progenitors of other carbon-rich giants with high binarity, including C-H, Ba, and CEMP-$s$ stars \citep[e.g.,][]{McClure1990, Hansen2016, Izzard2010}.  

\begin{figure}
\epsscale{1.3}
\plotone{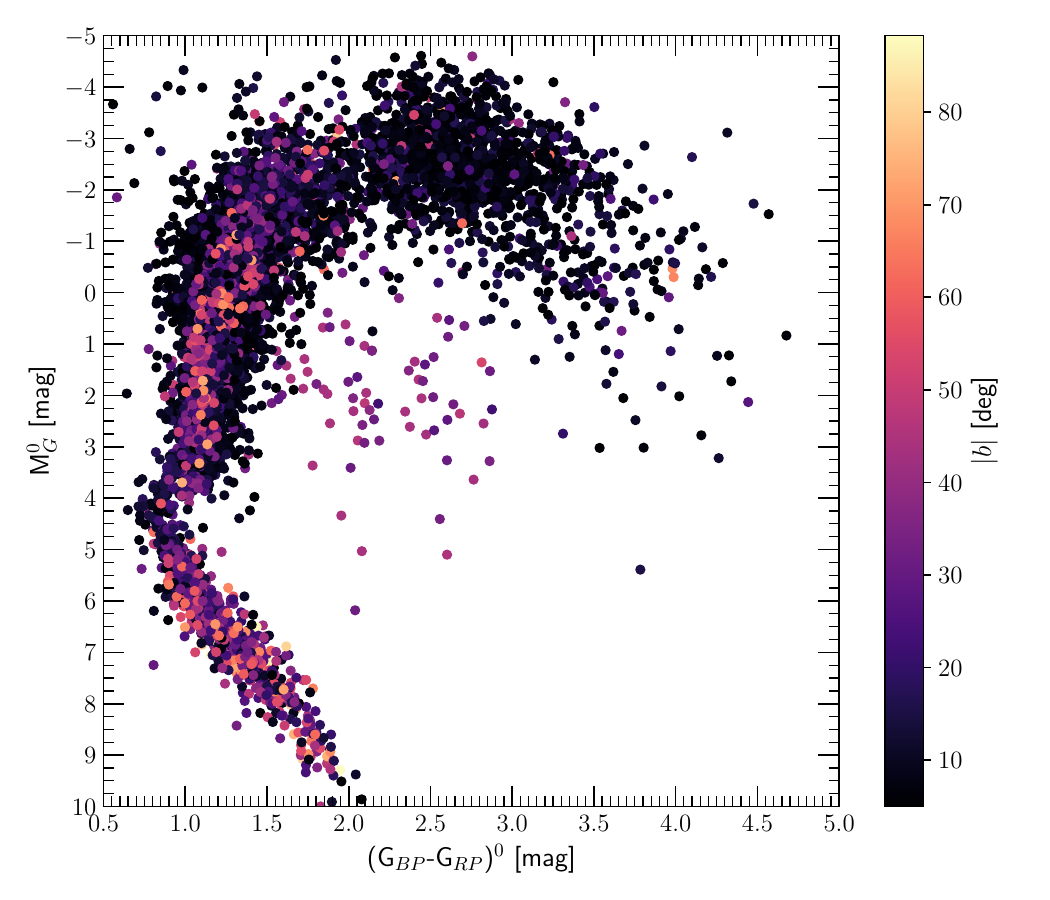}
\caption{High confidence carbon star sample in the \GDR3\, color-magnitude plane.  We plot the dereddened absolute $G$ magnitude versus dereddened color ($B_P-R_P$) for 9268 high confidence C stars (XG\_Prob\_C$>0.85$) with parallax/error $>3$ and Galactic latitude $|b|>5$. Point colors correspond to $|b|$. Carbon-rich stars cover a broad region of the diagram, including dwarfs, subgiants, red giants, and AGB stars. The majority of AGB stars are found at low Galactic latitude.}
\label{fig:CstarCMD}
\end{figure}

Carbon stars thus span nearly the full Hertzsprung-Russell diagram, from main sequence (dC) to giants and AGB stars, as was illustrated in the carbon star color–magnitude diagram shown by \citet[][Figure\,2]{Green2019}, and with a larger, more homogeneous sample here in Figure\,\ref{fig:CstarCMD}. While C-AGB stars are intrinsically bright, they are rare and expected to be $\sim10^3$ times less common than dC stars, given the relative durations of the AGB and main sequence phases \citep{Kool1995}. Nevertheless, early surveys with bright magnitude limits (e.g., \citealt{Stephenson1985, Sanduleak1988}) preferentially discovered the more luminous C giants. In contrast, dCs, being $10^{2-3} \times$ less luminous, are better sampled by deeper surveys such as SDSS and LAMOST \citep[e.g.,][]{Green2013, Li2024}, although these datasets suffer from strong selection effects and incomplete sky coverage.

As main sequence stars enriched by mass transfer, dCs span a wide range in absolute magnitude and color. Their $M_G$ values range from $\sim11$ for cool dCs ($\bprp \sim 2.2$) to $\sim4.5$ for warmer, more massive types ($\bprp \sim 0.7$). Representative spectra for dCs are shown in \citet[][Figure 1]{Roulston2020}, and their distribution in color–magnitude space is illustrated by \citet[][Figure 2]{Roulston2022}. Even hotter main sequence stars with C$>$O atmospheres may exist, but if they are too warm, molecular bands like C$_2$ and CN are likely undetectable.


Using machine learning and a training set of known, definitively-classified C stars from LAMOST DR8 \citep{Deng2012}, \citet{Roulston2025} produced an all-sky catalog of C star candidates using low-resolution (R\,$\sim 20 - 70$) spectra from {\em Gaia} Data Release 3 \citep[DR3;][]{GaiaDR3}.  The catalog includes C stars across a wide range of luminosities, but the catalog paper focused attention  on the dwarf carbon stars, their luminosity function and their space density. In the current work, we examine the C giants of the Milky Way, its globular clusters, and local group dwarf galaxies including the Magellanic Clouds.  Throughout the current work, we adopt their  catalog, but restrict our sample to the most reliable subset of C star candidates, those with a C star classification probability from the XGBoost algorithm (`XGProb\_C') greater than 0.85.  Such stars have a 95.5\% probability of being a true C star, based on inspection of intermediate resolution spectra (typical resolution of FWHM$\sim\,6-7$\AA), as described in \citet{Roulston2025}.  

\section{Carbon Stars Across the Color-Magnitude Diagram \label{sec:CMD}}

The carbon CMD in Figure\,\ref{fig:CstarCMD} is remarkable in highlighting how stars with strong molecular carbon bands span about 
15 mag in absolute magnitude $M_G$, or a factor of a million in luminosity.  Despite the spectroscopic similarities of stars across the carbon CMD, their origin stories differ widely.  Carbon main sequence stars (dCs, with $M_G \gtrsim 4$) have all inherited C-enhanced material from an AGB companion, which has since evolved to a white dwarf, as demonstrated by their binarity and mass function \citep{Roulston2019}.  Short-period variability and even X-ray emission demonstrate that many dC systems likely accreted significant mass and angular momentum, and some entered a common envelope phase with their C-AGB donor companion \citep{Roulston2021, Roulston2022}.  Dwarfs with C$>$O likely persist at absolute magnitudes $M_G\, \lesssim 4$, but these stars are too warm for the typical C star molecular bands. Detection of C$_2$ molecular bands becomes increasingly difficult for $T_{\rm eff}\gtrsim5000$\,K, and the Swan bands are generally absent by $T_{\rm eff}\gtrsim6000$\,K (e.g., \citealt{Keenan1993,Barnbaum1996,Bergeat2001}).  On the O-rich main sequence, these temperatures correspond to ($B_P-R_P$)$\lesssim 0.5$, or late-F to early-G spectral types \citep{Gray2009}.  The C-CMD in Figure\,\ref{fig:CstarCMD} lacks stars that blue, likely because our algorithm and choice of confidence level excludes stars with weak C$_2$ bands even if they are C-rich.  



\begin{figure*}
\centering
\gridline{\fig{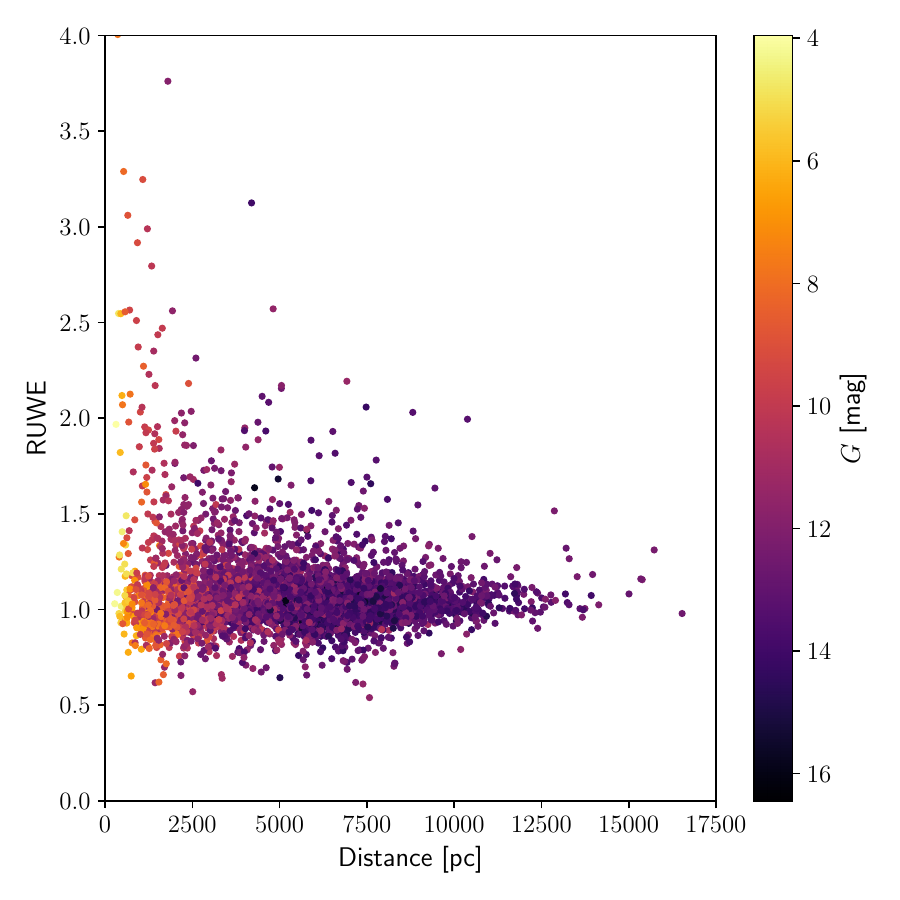}{0.5\textwidth}{(a)}
          \fig{RUWE_loghisto.pdf}{0.5\textwidth}{(b)}}

\caption{(a) RUWE, an astrometric indicator of binarity, is plotted vs. distance from \citet{Bailer-Jones2021} for our MW C star sample. Point colors indicate apparent $G$ mag.  Binaries are more easily detected via RUWE, as expected, for small distances. 
(b) The normalized RUWE histograms, on a log scale, limited to stars with distance $<$3\,kpc, for 418 C giants ($-4<M^0_J<0$; blue) and 395 AGB stars ($M^0_J < -5$; red).  Large RUWE values indicating binarity are much more common for C-RGB stars, confirming expectations that the C-rich nature of the lower luminosity red giants is extrinsic.  }  
\label{fig:RUWE}
\end{figure*}

The C-AGB stars have absolute magnitudes \hbox{$M_G \lesssim -1$}.  There is no obvious break between lower luminosity carbon red giants (C-RGB hereafter) and C-AGB stars in the $G$-band CMD of Figure\,\ref{fig:CstarCMD}, and the choice of a separatrix in any CMD depends on the desired balance between completeness and purity.  The origin of C-RGB stars has been discussed for decades, typically as CH or R-type giants.  C-RGB stars, as they do not dredge up their own carbon-enhanced material must be evolved post-mass-transfer systems with a WD companion, probably evolved dC stars.  With our sample, we can test this by comparing the distribution of the RUWE for C-RGB vs. C-AGB stars.  Figure\,\ref{fig:RUWE} shows indeed that large RUWE values indicating binarity are much more common for C-RGB stars, confirming the expectations that their C-rich nature is often extrinsic.  In this figure, we contrast luminous C-AGB stars (defined here by $M^0_J<-5$) with C-RGB stars ($M^0_J>-4$) out to a distance of 3\,kpc, within which astrometric motions caused by binarity can be detected with the \GDR3\, RUWE metric. For 418 such C-AGB stars, the median RUWE is 0.99, compared to 1.11 for the 395 C-RGB stars. The fraction of C-AGB stars with RUWE$>$1.4 is 3.6\%, compared to 22.8\% for C-RGB stars.  These RUWE distributions differ significantly according to a two-sample Kolmogorov-Smirnov test ($D$=0.32, $p<0.001$).

Early-R (R-hot) carbon stars are likely included in the C-RGB sample above.  These are now widely interpreted as single red-clump giants whose carbon enrichment results from non-canonical mixing during the core-helium flash, rather than from third dredge-up or binary mass transfer \citep{Knapp2001,Dominy1984}. Their luminosities, lack of $s$-process enhancement, near-solar metallicities, and low binary fraction strongly support a helium-flash origin, possibly aided by rotation- or merger-induced mixing \citep{Izzard2007,Zhang2020}.  Without the early-R stars in our C-RGB sample, the RUWE distributions discussed above would likely be even stronger.

Other stars that have experienced mass transfer from a former TP-AGB companion include the Ba, extrinsic S, CH, and CEMP-$s$ stars \citep{Wallerstein1998, Jorissen1998, Jorissen2016}. Their differing observational classifications arise primarily from metallicity- and temperature-dependent chemistry; at solar metallicity, $s$-process enrichment produces Ba and S stars without forming C stars, while at low metallicity the same enrichment drives C/O$>$1, producing CH \citep{Vanture1992} and CEMP-$s$ stars \citep{Beers2005,Lucatello2005}.  The Ba and CEMP-$s$ stars are unlikely to be well-represented in our C star catalog, as they often have $T_{\rm eff} \gtrsim 5000$\,K where, although CH bands may persist, the C$_2$ Swan bands become quite weak. This is particularly true for Ba stars, which have metallicities near solar (e.g., \citealt{deCastro2016, Escorza2019}), so that CO leaves little carbon to form C$_2$ or CN.  By contrast, some cool CEMP-$s$ giants are known to have strong C$_2$ and CN bands (e.g., \citealt{Beers2005,Aoki2007,Masseron2010}).

\section{C Stars in Globular Clusters \label{sec:GCs}}

To find C stars associated with globular clusters, we use the {\em Gaia} EDR3 catalog of \citet{Vasiliev2021}, which provides coordinates, mean proper motion and parallax for 170 clusters.  We positionally match 70 C stars within four times the listed Plummer scale radius $R_0$ for all clusters\footnote{At $4R_0$, the surface density of stars is nominally about 0.1\% the central density \citep{Plummer1911}.}, finding C star matches for 34 of the clusters. For most clusters, we find a single associated C star to our magnitude limit of $G=16.5$\,mag.  For 47~Tuc, we find 11, for $\omega$~Cen we find 9, and 4 for NGC\,362.  A list of the 70 C stars matched to globular clusters is provided in Table\,\ref{tab:CGCmatched}.

\begin{figure}
\includegraphics[width=\linewidth]{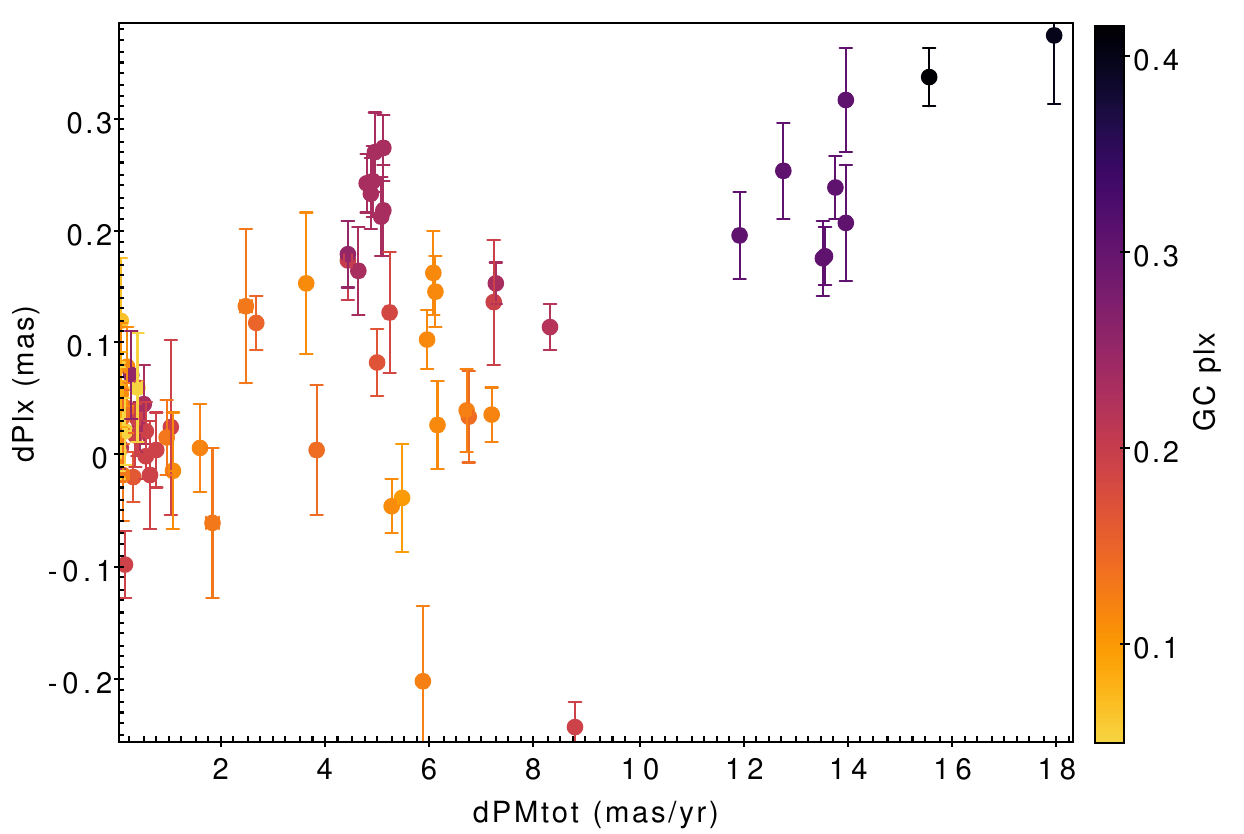}
\caption{For 70 carbon stars matched within four Plummer scale radii $R_0$ to 34 globular cluster (GC) central positions, we plot the difference dPlx between the mean GC and C star parallax, versus the difference dPMtot between the mean GC and C star proper motion.  The point colors represent the mean GC parallax values. The least trustworthy matches are the 9 C stars with dPMtot$>10$\,milliarcsec/yr, of which 7 are associated with M\,22. These 9 also are the matched C stars closest to the Sun.}
\label{fig:GC_dPlx_dPMtot}
\end{figure}

Cluster membership is highly likely for most of the matched C stars.  The majority of C stars are not only positionally associated, but show relatively small deviations in parallax or proper motion from the positionally-matched cluster.  As evident in Figure\,\ref{fig:GC_dPlx_dPMtot}, some of the C stars show larger discrepancies in parallax and proper motion, especially for 7 C stars associated with NGC 6656 (M\,22).  M22 may be experiencing mild tidal mass loss \citep{Kunder2014}, but there is currently no strong {\em Gaia}-era evidence for large-scale, well-defined tidal tails indicative of severe ongoing disruption \citep{Sollima2020}.  

The absolute magnitudes $M_G$ of the C giants we find in globular clusters are listed in  Table\,\ref{tab:CGCmatched}, calculated at the distance of the cluster.  All of those $M_G$ values are plausible, i.e., brighter than the listed $G<16.5$ magnitude limit $M_G^{lim}$ of our sample at the distance of each matched cluster.  All the C giants are in the subgiant to red giant luminosity range $-4\lesssim M_G \lesssim 4$.  We cannot detect most dC stars $M_G\gtrsim 5$ at these distances. We find no C-AGB stars in any of the clusters, which is not unexpected, as the vast majority (all but one, Pal 12) of the matched globular clusters are at least 10\,Gyr old \citet{VandenBerg2013}, which corresponds to a turnoff mass of about $0.8\,\Msun$  (e.g., \citealt{Dotter2008}), whereas the third dredge up in the TP-AGB phase that produces C-AGB stars does not occur below about $1.5\,\Msun$ (e.g., \citealt{Marigo2017}).

\startlongtable
\begin{deluxetable*}{rrrrlr}
\label{tab:CGCmatched}
\tablecaption{Globular Cluster C Stars\label{tab:gc_sample}}
\tablewidth{0pt}
\tablehead{
\colhead{R.A. (deg)} & \colhead{Dec (deg)} & \colhead{source\_id} & \colhead{$M_G$} & \colhead{Globular Cluster} & \colhead{$M^{lim}_G$} 
}
\startdata
    5.27044 & -72.25643 & 468961294301162905620 &   2.84 & NGC 104, 47 Tuc &   3.33 \\ 
    5.43803 & -72.45910 & 468954855713036288020 &   2.92 & \ldots\ &   \ldots\ \\ 
    5.86690 & -72.44351 & 468956824526115737620 &   2.53 & \ldots\ &  \ldots\  \\ 
    5.93628 & -72.02915 & 468963916376527667220 &   1.14 & \ldots\ &  \ldots\ \\ 
    6.01922 & -71.73678 & 468983841587462144020 &   1.51 & \ldots\ &  \ldots\ \\ 
    6.27110 & -72.44963 & 468955721578501836820 &   2.55 & \ldots\ &  \ldots\ \\ 
    6.67845 & -72.10566 & 468962947003377868820 &   2.82 & \ldots\ &  \ldots\ \\ 
    6.69681 & -72.27619 & 468957697263439872020 &   3.01 & \ldots\ &  \ldots\  \\ 
    7.10142 & -71.75753 & 468965297638160384020 &   2.27 & \ldots\ &  \ldots\ \\ 
    7.33604 & -72.04724 & 468958391330138726420 &   2.97 & \ldots\ &   \ldots\ \\ 
   15.66642 & -70.77454 & 469088765921045913620 &   1.59 & NGC 362  &   1.78 \\ 
   15.95612 & -70.73015 & 469088944591681331220 &   0.99 & \ldots\ & \ldots\   \\ 
   16.21520 & -70.76671 & 469084181901908889620 &   1.50 & \ldots\  & \ldots\   \\ 
   16.23027 & -70.72531 & 469088893052073062420 &   0.73 & \ldots\  & \ldots\   \\ 
  189.66859 & -51.14630 & 607898085401766707220 &   0.53 & Rup 106  &   0.63 \\ 
  193.31447 & -67.10425 & 585816345561444147220 &  -1.03 & BH 140   & 3.20   \\ 
  193.38196 & -67.27304 & 585811085519941427220 &  -0.58 & \ldots\  &  \ldots\  \\ 
  194.86278 & -70.84877 & 584379899335170662420 &   0.19 & NGC 4833  &   2.57 \\ 
  201.41905 & -47.35807 & 608371747711170252820 &   1.79 & NGC 5139, $\omega$ Cen &   2.93 \\ 
  201.45823 & -47.45505 & 608371346132611584020 &   2.43 & \ldots\  & \ldots\ \\ 
  201.50660 & -47.55169 & 608370026719663104020 &  -2.59 & (1) & \ldots\ \\ 
  201.56007 & -47.46819 & 608370116052510105620 &  -0.83 & \ldots\ & \ldots\ \\ 
  201.61726 & -47.39460 & 608371462957058355220 &  -2.09 & \ldots\ & \ldots\ \\ 
  201.66713 & -47.48097 & 608370163726356684820 &  -1.39 & \ldots\ & \ldots\ \\ 
  201.71755 & -47.43541 & 608370215265970176020 &  -1.74 & \ldots\ & \ldots\ \\ 
  201.92010 & -47.71999 & 608350998294593228820 &  -1.93 & \ldots\ & \ldots\ \\ 
  202.16639 & -47.44258 & 608389250990329139220 &  -2.49 & \ldots\ &  \ldots\ \\ 
  229.69109 &   2.16233 & 442157527396489369620 &   0.19 & NGC 5904, M 5 &   2.25 \\ 
  236.44691 & -37.79147 & 600955191741875200020 &   1.44 & NGC 5986  &   1.73 \\ 
  236.46518 & -37.78422 & 600955195177850265620 &  -0.23 & \ldots\  & \ldots\ \\ 
  246.36292 & -72.24341 & 580647009253633536020 &   0.98 & NGC 6101  &   1.12 \\ 
  255.72791 & -30.03039 & 602938370565090099220 &   2.49 & NGC 6266, M 62 &   2.84 \\ 
  255.80190 & -26.21275 & 411191682500236032020 &   1.39 & NGC 6273, M 19 &   2.26 \\ 
  256.26966 & -22.80234 & 411393958275509760020 &  -0.21 & NGC 6287  &   2.37 \\ 
  258.52834 & -29.51565 & 410735598205347072020 &   0.60 & NGC 6304  &   2.64 \\ 
  259.08935 & -28.08015 & 410774716330991820820 &   1.38 & NGC 6316  &   1.77 \\ 
  259.15014 & -28.17763 & 410774132656267417620 &   1.66 & NGC 6316  &   1.77 \\ 
  260.21918 & -19.64508 & 412205493632620953620 &   2.09 & NGC 6342  &   2.25 \\ 
  262.67282 & -39.90072 & 596016106488283340820 &   0.51 & FSR 1758  &   1.93 \\ 
  262.82187 & -39.66216 & 596167196563672883220 &  -3.62 & \ldots\  & \ldots\ \\ 
  262.91783 & -67.06894 & 581308228059684966420 &   1.24 & NGC 6362  &   2.17 \\ 
  264.97061 & -53.96904 & 592136830173259468820 &   2.88 & NGC 6397  &   4.60 \\ 
  266.21555 &   3.18224 & 437629803796739891220 &   0.03 & NGC 6426  &   0.04 \\ 
  266.22940 &   3.16941 & 437629783180531302420 &  -1.90 & \ldots\  & \ldots\ \\ 
  267.71379 & -34.57870 & 404158264235735705620 &   0.89 & NGC 6453  &   1.46 \\ 
  269.78765 & -44.28988 & 595605307891451289620 &  -0.49 & NGC 6496  &   1.88 \\ 
  271.75961 & -25.12776 & 406568830093118617620 &   3.55 & NGC 6544  &   4.50 \\ 
  272.00244 & -43.70593 & 672404209099330355220 &  -1.13 & NGC 6541  &   2.31 \\ 
  272.39262 & -26.01770 & 406482340194887884820 &   1.71 & NGC 6553  &   2.94 \\ 
  273.51217 & -31.89120 & 404888347269021798420 &  -0.56 & NGC 6569  &   1.75 \\ 
  275.83643 & -30.43746 & 404645873591137996820 &   1.85 & NGC 6624  &   2.07 \\ 
  275.88694 & -30.45497 & 404641162871241676820 &   1.98 & \ldots\  & \ldots\ \\ 
  275.95993 & -30.33964 & 404646086621550796820 &   1.25 & \ldots\  & \ldots\ \\ 
  276.27464 & -24.83927 & 407719304518987878420 &   1.19 & NGC 6626, M 28 &   3.00 \\ 
  278.72893 & -24.19867 & 407748327755138457620 &   1.31 & NGC 6656, M 22 &   3.93 \\ 
  278.87824 & -23.50173 & 407766026890273433620 &   3.12 & \ldots\ & \ldots\ \\ 
  278.91563 & -32.92701 & 673661501078522880020 &   0.73 & NGC 6652  &   1.88 \\ 
  279.00953 & -24.27095 & 407672905131601612820 &   3.20 & NGC 6656, M 22 &   3.93 \\ 
  279.06808 & -23.44620 & 407761982315697920020 &   2.86 & \ldots\ & \ldots\ \\ 
  279.09746 & -24.21878 & 407673523606972006420 &   3.74 & \ldots\ &  \ldots\ \\ 
  279.22840 & -24.06000 & 407673839278329190420 &   1.91 & \ldots\ & \ldots\ \\ 
  279.30605 & -23.86926 & 407683910117884211220 &   2.69 & \ldots\ &  \ldots\ \\ 
  283.60117 & -30.41959 & 676117926877060608020 &  -1.75 & NGC 6715 M 54 &   0.12 \\ 
  283.71714 & -30.40636 & 676043029519789875220 &  -0.69 & \ldots\ & \ldots\ \\ 
  287.37194 & -59.70000 & 663258227854378496020 &   2.25 & NGC 6752  &   3.52 \\ 
  287.77357 &   0.97650 & 426731770965054976020 &   0.92 & NGC 6760  &   2.10 \\ 
  298.50394 &  18.84821 & 182161517210553856020 &   2.19 & NGC 6838 M 71 &   3.50 \\ 
  313.38090 & -12.53379 & 688946891108147200020 &  -0.26 & NGC 6981 M 72 &   1.12 \\ 
  326.70877 & -21.27429 & 681795079929458585620 &  -0.86 & Pal 12  &  -0.01  
\enddata
\tablecomments{This table lists 70 high confidence C stars discovered within four Plummer scale radii of 34 globular clusters. $M_G$ is the (un-dereddened) absolute magnitude of each C star at the distance of its associated cluster. $M^{lim}_G$ is the faintest absolute magnitude (corresponding to our survey limit of $G=16.5$\,mag) detectable for each cluster, given its distance.}
\footnote{(1) This is a foreground giant at 3190\,pc distance, and is detected in X-rays by {\em Chandra}.}
\end{deluxetable*}

Several of these C stars associated with NGC\,362 (4 C stars) and 47 Tuc (7 C stars) were found by \cite{Morgan1995} in their objective prism plate survey of the outskirts of the SMC.  Indeed, these two globular clusters are near on the sky, but not physically related to the SMC. Seven of nine C stars we find in $\omega$\,Cen are previously known \citep{vanLoon2007}. One in FSR\,1758 \citep{Westerlund1971} and one in M54  \citep{McDonald2012} are also previously known; the latter may alternatively be associated with the Sgr dwarf spheroidal galaxy.  


\section{C Stars in Local Group Dwarf Galaxies \label{sec:dGals}}

For matching C stars to Local Group Dwarf Galaxies, we require that the \GDR3\, renormalized unit weight error not exceed RUWE $=1.4$, above which binarity\footnote{We will explore C star variability and binarity in future work.} may strongly affect the astrometric solution \citep{Castro-Ginard2024}.  The latter cut reduces the number of C stars in the LMC, SMC, the Sagittarius dwarf spheroidal, and the Milky Way  by 3.9\%, 1.6\%,	1.0\%, and 7.8\%, respectively.  Further position, parallax and proper motion cuts described below identify high confidence C star candidates in these galaxies.

\subsection{C Stars Associated With the Large Magellanic Cloud \label{sec:LMC}}

\begin{figure}
\includegraphics[width=\linewidth]{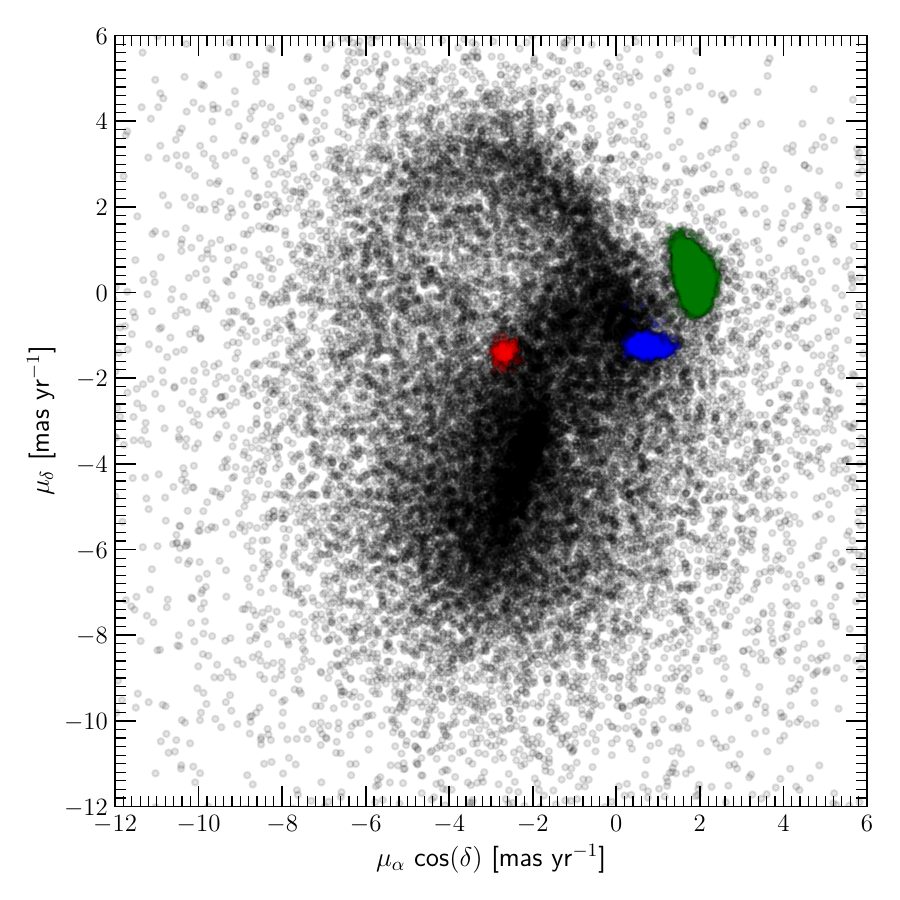}
\caption{Proper motion in R.A. and Dec from \GDR3\, for C stars spatially associated with the LMC (green), SMC (blue) and Sgr (red) dwarf galaxies, along with all the C stars in our sample (grey).  We use this proper motion information, along with parallax constraints to identify high confidence association with these galaxies.}
\label{fig:XMCpms} 
\end{figure}

 Proper motions for C giants in the Magellanic Clouds are particularly helpful to distinguish them from C giants in the Milky Way.   The mean proper motions for the LMC and SMC have been published by \citet{Kallivayalil2013}, based on long-baseline {\em HST} imaging epochs.  Subsequent, detailed analyses based on {\em Gaia} data are published by a variety of authors (e.g., \citealt{Patel2020} and references therein). We plot in Figure\,\ref{fig:XMCpms} proper motions from \GDR3\, for C stars spatially associated with the LMC, SMC and Sgr.  We use this proper motion information along with parallax constraints to identify high confidence association with these galaxies.  
 
The distance of the LMC is about 50\,kpc.  \citet{Pietrzynski2019} report a distance modulus $\mu_{LMC}=18.477\pm 0.026$, corresponding to a parallax of 0.0202\,mas. From our parent C star catalog, we define LMC C stars to be those meeting the following criteria: (1) within 11$\deg$ of the LMC center, as listed in the NASA/IPAC Extragalactic Database (2) parallax within 2$\sigma$ of 0.02 mas (3) R.A. proper motion between 1.2 and 2.5 mas/yr and (4) Dec proper motion between $-0.6$ and 1.6 mas/yr.  These further cuts yield 6738 high confidence C stars associated with the LMC.
Given the distance of the LMC, our $G<16.5$ parent catalog includes only C giants with $M_G<-2.27$. 
To calculate absolute magnitudes of our C giants, we adopt a mean LMC color excess of $E(B-V)=0.127$ from \citep{Gorski2020}.

\subsection{C Stars Associated With the Small Magellanic Cloud \label{sec:SMC}}

The distance of the SMC is about 62\,kpc.  \citet{Graczyk2020} report a distance modulus $\mu_{LMC}=18.977\pm 0.032$, or parallax 0.016\,mas. 

From our parent C star catalog, we define LMC C stars to be those meeting the following criteria:
(1) within 9$\deg$ of the SMC center, as listed in the NASA/IPAC Extragalactic Database (2) parallax within 2$\sigma$ of 0.016 mas (3) R.A. proper motion between 0.15 and 1.5 mas/yr and (4) Dec proper motion between $-1.6$ and 0.8 mas/yr.
These further cuts yield 2114 high confidence C stars associated with the SMC.
Given the distance of the SMC, our $G<16.5$ parent catalog includes only C giants with $M_G<-2.7$. 
To calculate absolute magnitudes of our C giants, we adopt a mean SMC color excess of $E(B-V)=0.084$ from \citep{Gorski2020}.

\subsection{C Stars Associated With Other Local Group Galaxies \label{sec:LG}}

To find which of the C stars in our catalog are not part of the Milky Way but might rather associated with Local Group dwarf galaxies, we use the latest (2021) public update of observed properties of 144 dwarf galaxies in and around the Local Group, originally compiled and described in \citet{McConnachie2012} and available via the CADC\footnote{https://www.cadc-ccda.hia-iha.nrc-cnrc.gc.ca/en/community/nearby/}.  Since the apparent mag limit of our C star survey is 16.5, and since the most luminous C giants have $M_G= -5$ mag, we would not expect to detect any C giants in galaxies with a distance modulus larger than 21.2 (200\,kpc). That cut leaves 55 galaxies wherein we might detect C stars.  The nearest galaxy in the catalog is Canis Major at 7.2\,kpc, corresponding to a parallax of 0.14 milliarcsec.  

Next, we should only bother to match high confidence C stars from our catalog that have parallaxes consistent with possible membership in the galaxies.  We conservatively consider only stars from our catalog with XG\_Prob\_C$>0.85$, and \GDR3\, parallax $<0.12$mas or parallax $<$ parallax\_error, which leaves 23,811 C stars for initial matching to Local Group dwarf galaxies.

We first perform matching on the celestial sphere, including any C star within an ellipse at the cataloged position angle, with major axis $a$ twice the half-light radius listed in the galaxy catalog, and minor axis $b=a(1-e)$ using the listed ellipticity. We thus find C stars spatially associated with 7 dwarf galaxies besides the Magellanic Clouds - Crater\,II, Carina, Reticulum\,II, Sculptor, Antlia\,II, and the Sagittarius dwarf spheroidal.  However, we must ensure not only spatial association on the sky, but proper motions, parallaxes and (when available from \GDR3) velocities consistent with membership.  All such data are consistent with membership for one C giant each in Carina, Reticulum\,II and Sculptor. However, we find many C stars convincingly associated with the Sagittarius Dwarf Spheroidal.

\subsubsection{The Sagittarius Dwarf Spheroidal}\label{sec:Sgr}
The Sagittarius dwarf spheroidal galaxy (Sgr hereafter), first discovered by \citet{Ibata1994}, is the nearest satellite galaxy of the Milky Way, and the brightest dwarf spheroidal galaxy in the Local Group. At a distance of about 25\,kpc \citep{Toguchi2025} Sgr is interacting - and probably merging - with the Milky Way galaxy.  Using \emph{HST} observations of stars against background galaxies, \citet{Massari2013} found the proper motion of Sgr to be $(\mu_{\alpha} \cos \delta, \mu_{\delta}) = (-2.54 \pm 0.18,\,-1.19 \pm 0.16)~\mathrm{mas\ yr^{-1}}$, while most Galactic stars in the same region have $(\mu_{\alpha} \cos \delta, \mu_{\delta}) = (-1.21 \pm 0.27,\,-4.39 \pm 0.26)~\mathrm{mas\ yr^{-1}}$.

\begin{figure*}
\epsscale{1.0}
\plottwo{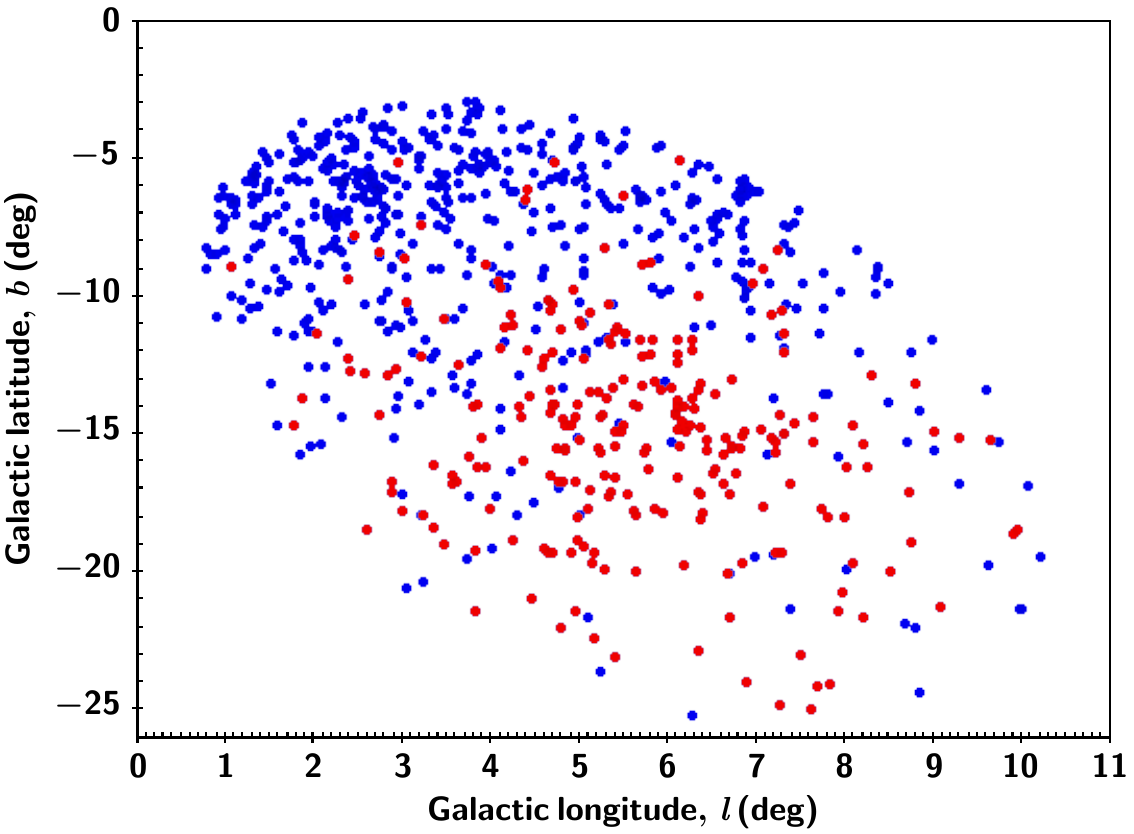}{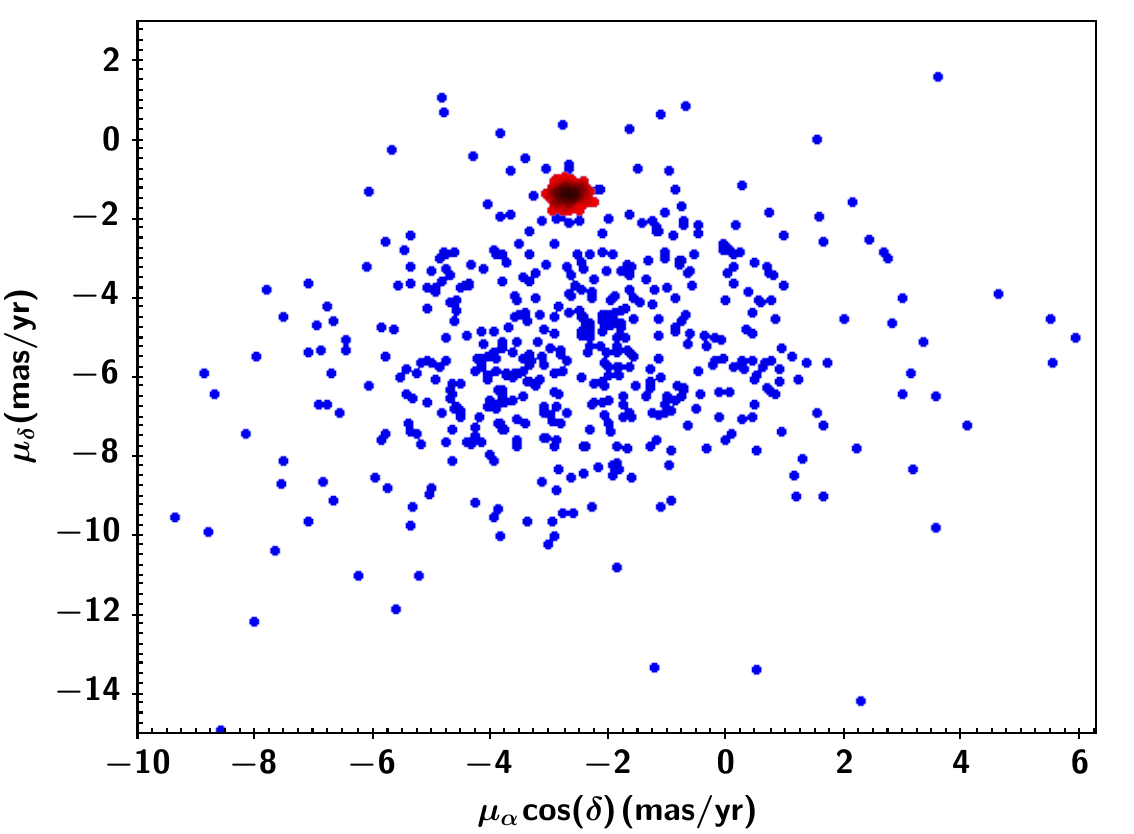}
\caption{LEFT: Galactic latitude and longitude of high confidence carbon stars in the region considered for membership in the Sagittarius dwarf spheroidal galaxy. RIGHT: RA and Dec proper motion of all stars within this sky region. A dense clump is obvious near the mean proper motion values for Sgr measured using $HST$ images by \citet{Massari2013}. For both plots, blue points show the full C giant sample within the elliptical sky region of Sgr, while red points show those C stars within our selected proper motion range. }
\label{fig:SgrPM}
\end{figure*}

We count as potential members of Sgr those C stars within the ellipse of major axis twice the half-light radius ($r_h=5.7$ arcmin), with position angle 102$^{\circ}$ and ellipticity $e=0.64$, as listed in  \citet{Massari2013}. This region of sky overlaps the Galactic plane towards the Galactic center, extending to $b=-3$ at $l\sim 4$, such that the majority of the 765 C stars we detect in this sky region are probably not Sgr members.  Since the proper motion of Sgr offers strong contrast with the Galactic C stars (see Figure\,\ref{fig:SgrPM}, we restrict the list of C stars positionally-matched to Sgr to within a proper motion range between $-2.2 > (\mu_{\alpha} \cos \delta) > -3.1$ and $-1.9<\mu_{\delta}<-0.9$.  Of the 246 C stars thus designated as Sgr members, 99 have radial velocity measurements in \GDR3.  The mean and 1$\sigma$ dispersion of these velocities is 140$\pm 16\kms$, in excellent accord with Sgr measured core mean velocity of 140$\pm 2\kms$, adding to our confidence in our selection methods.  We further restrict the parallax measurements to be within 2$\sigma$ of 0.038 mas \citep{Toguchi2025}, which yields 219 high-confidence C giants associated with Sgr.


There are several papers in the literature listing C stars in Sgr and its streams. 
The largest compilations are in \citet{McDonald2012}, \citet{Whitelock1999}, \citet{Battinelli2013} and \citet{Lagadec2009}, which list 67 unique C stars (11 are listed in two papers rather than just one).  Of those, 66 have a \GDR3\, match within 2arcsec, of which 59 are brighter than our $G=16.5$ C star survey magnitude limit.  Three objects from \citet{Battinelli2013} and one from \citet{McDonald2012} have \GDR3\, parallax and/or proper motions inconsistent with Sgr membership: Sgr-4, Sgr-6 and Sgr-9. 
Of 55 published C stars likely to be members of Sgr that we might expect to detect, only 35 are included in our list of high confidence C stars in Sgr.  There is no obvious difference in photometric or astrometric properties between the 35 matched and the 20 unmatched, reflecting the need for further study to enhance survey completeness for C stars.  Our criteria may not include some stars with weak C$_2$ and CN bands that were selected by other studies.  On the other hand, our selection nearly quadruples the catalog of {\em bona fide} Sgr C giants brighter than $G=16.5$\,mag.



Given the distance of Sgr, the apparent magnitude limit of our parent catalog includes only C giants with $M_G<-0.84$. 
To calculate absolute magnitudes of Sgr C giants, we adopt a mean reddening value of {\it E(B-V)}$=0.15$, 
found using fits to the {\em HST} color-magnitude diagram (CMD) of stars in Sgr  \citep{Siegel2007}.  
This value remains relatively constant across the galaxy, according to \citet{Vitali2025}.

\section{Carbon Giants in the Milky Way \label{sec:MW}}

We now consider C giants in the Milky Way.  We include only those XG\_Prob\_C$>0.85$ C stars with significant parallax ($>3\times$parallax\_error), and a dereddened absolute magnitude $M_G<0$.   To avoid effects of crowding and high extinction, we remove from consideration C stars close to the Galactic plane with $|b| < 10^\circ$.  We eliminate C stars associated as described above with Globular Clusters, the LMC, SMC and Sgr.  This leaves 2238 Milky Way C giants for study.   

\begin{figure*}
\gridline{\fig{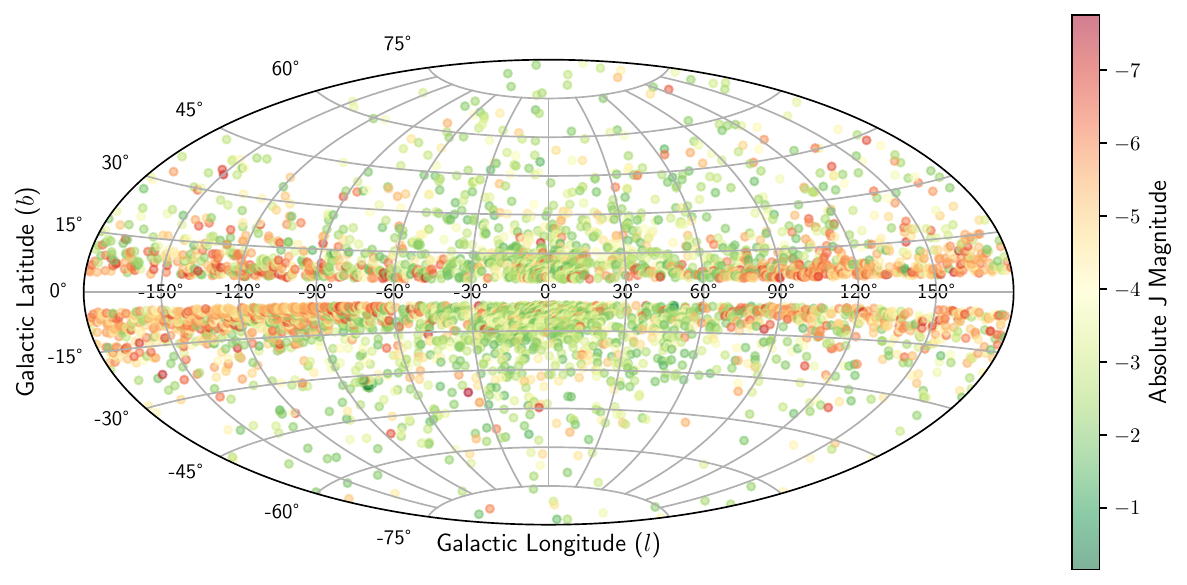}{0.5\textwidth}{} \fig{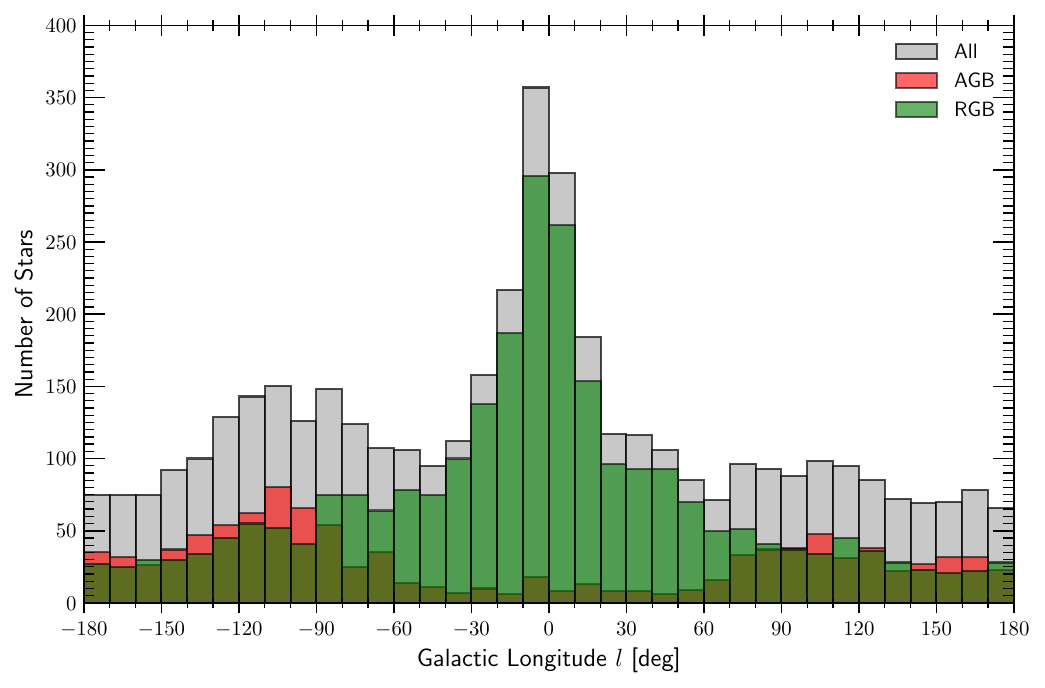}{0.5\textwidth}{}}
\caption{High confidence carbon giants in the Milky Way. LEFT: Aitoff projection in Galactic longitude $l$ and latitude $b$ for high confidence carbon giants in the Milky Way, with dereddened absolute mag $M^0_J$ as a colorbar.  RIGHT: Histogram showing the distribution in Galactic longitude $l$ of all C giants (grey), RGB (green) and AGB (red) C stars, as defined in \S\,\ref{sec:spacedist}.  C-AGB stars ($M^0_J \lesssim -5$) are rare towards the Galactic center in both plots.}
\label{fig:lGalbGal}
\end{figure*}

\begin{figure*}
\gridline{\fig{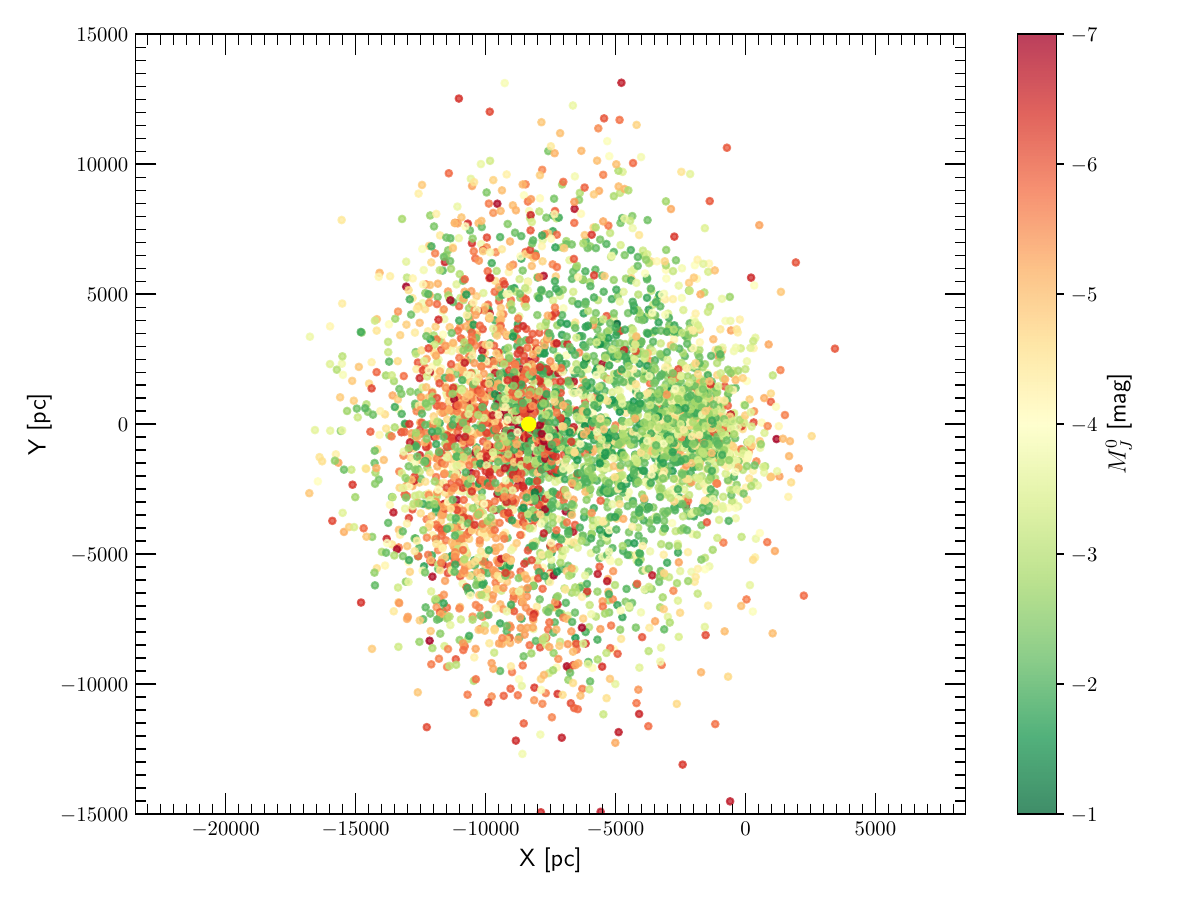}{0.5\textwidth}{} \fig{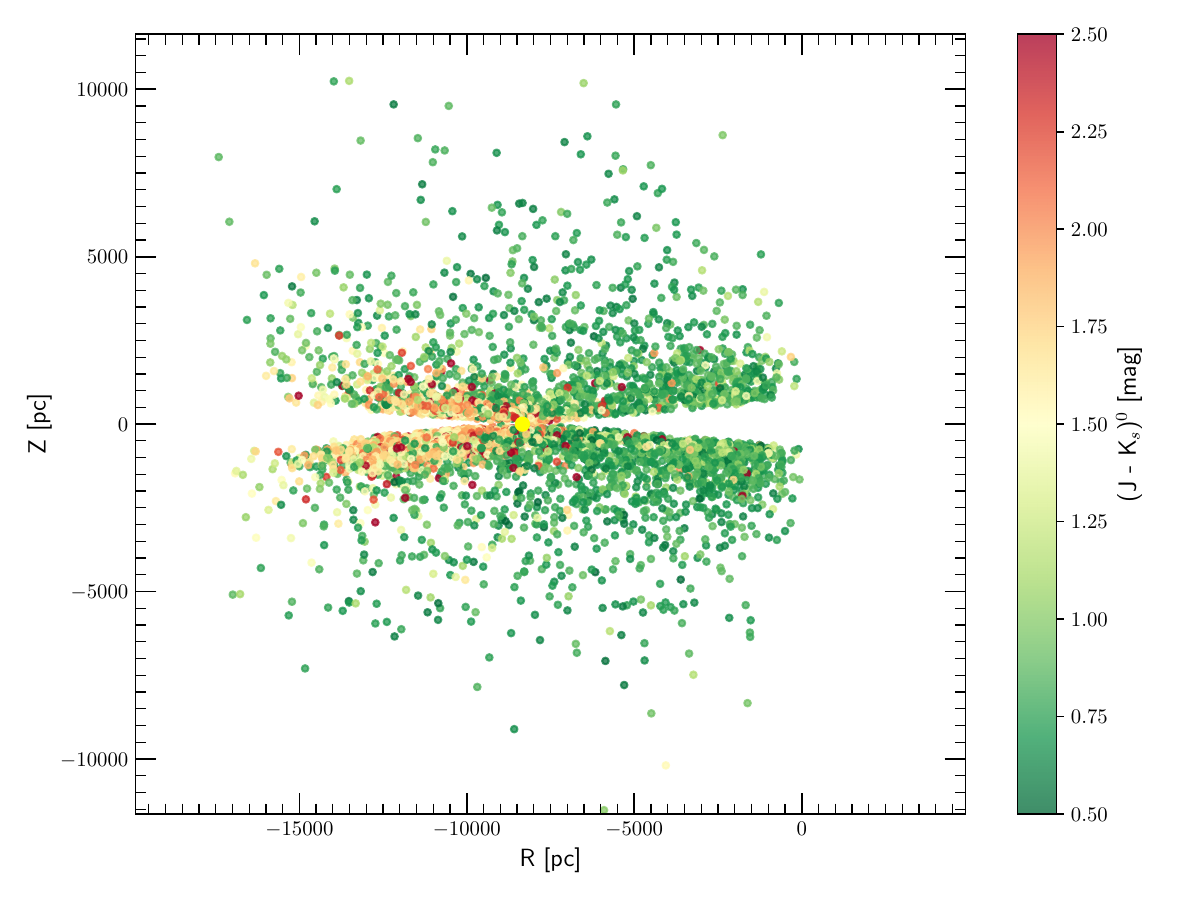}{0.5\textwidth}{}}
\caption{High confidence carbon giants in the Milky Way. LEFT: Y vs. X in pc relative to the Galactic center, with dereddened absolute mag $M^0_J$ as a colorbar. RIGHT: Z vs. R in pc relative to the Galactic center, with color ($J-K$)$^0$ as a colorbar. The location of the Sun is shown in each plot as a large yellow dot. C-RGB stars, likely extrinsically C-rich, dominate towards the Galactic center and in the halo, whereas C-AGB stars dominate the disk, away from the Galactic center.}
\label{fig:XYRZ}
\end{figure*}

We corrected for extinction as in \citep{Roulston2025} by using the \texttt{MWDUST} code \citep{Bovy2016}, using the `\texttt{combined19}' option which combines the maps of \citet{Drimmel2003}, \citet{Marshall2006}, and \citet{Green2019_dustmaps}.  

Figure\,\ref{fig:lGalbGal} and Figure\,\ref{fig:XYRZ} show the distribution of carbon giants in Galactic coordinates.\footnote{We adopt a Galactocentric coordinate system whereby $X=r\,cos(l)\,cos(b) - R_{\odot}$ and $Y=r\,sin(l)\,cos(b)$ and $Z=r\,sin(b) + Z_{\odot}$.  Here, $r$ is  median heliocentric stellar distance \texttt{r\_med\_geo} from \citealt{Bailer-Jones2021}, $R_{\odot} = $8178\,pc from \citealt{GRAVITYCollaboration2019}, and $Z_{\odot} =$ 25\,pc from \citealt{Reid2014}.}  A striking feature of these distributions is the pronounced deficit of C-AGB stars toward the Galactic center. This trend is likely driven, at least in part, by the strong radial metallicity gradient in the Galactic disk, with metallicity increasing inward at a rate of approximately $d[\mathrm{Fe}/\mathrm{H}]/dR \simeq -0.06,\mathrm{dex\,kpc^{-1}}$ \citep{Imig2023}. At the near-solar and super-solar metallicities characteristic of the inner disk, the higher initial oxygen abundance makes it increasingly difficult for third dredge-up episodes to raise the surface abundance above ${\rm C/O}>1$ \citep{Marigo2017}. Consistent with this picture, intrinsic TP-AGB carbon stars are known to be rare in the old, metal-rich Galactic bulge \citep{Blanco1989,Azzopardi1991}, which extends to a radius of only $\sim2$--3\,kpc from the Galactic center \citep{Portail2017}. Our results suggest that this paucity of C-AGB stars extends well beyond the bulge and into the inner disk.

A second contributing factor may be the radial age gradient of the Galactic disk. Stellar populations become progressively older toward smaller Galactocentric radii, with typical age gradients of order $d\mathrm{Age}/dR \sim -0.3$ to $-0.8\,\mathrm{Gyr\, kpc^{-1}}$ \citep{Johnson2025}. Because luminous carbon-rich AGB stars arise primarily from intermediate-age progenitors, their numbers are expected to decline in regions dominated by older stellar populations. Together, the metallicity and age gradients provide a natural explanation for the observed scarcity of C-AGB stars within approximately 7\,kpc of the Galactic center.

\section{Near Infrared Luminosity Functions \label{sec:LFs}}

Carbon star LFs (CSLFs) from well-understood, homogeneously-selected samples are important for refining models of AGB evolution to properly account for mass, age, and metallicity.  As an example, the use of variable molecular opacities to model the evolutionary properties of luminous AGB stars has been clearly demonstrated as necessary to reproduce the near-IR CMD for the LMC \citep{Marigo2003,Marigo2008}.  

The tip of the red giant branch (TRGB) can provide extragalactic distances to a few percent.  Near-infrared photometry is especially promising, since it is typically $1-2$ mag brighter than optical $I$ band, and is less affected by extinction. Accurate TRGB-based distances could extend to well beyond 20\,Mpc e.g., with JWST \citep{Max2024},  but the TRGB magnitude is best measured if the AGB C star contribution can be removed.  However, the direct use of C-AGB stars in the near-infrared is now receiving considerable attention as a distance estimator \citep{Ripoche2020,Lee2025}.  The absolute $J$ magnitude $M^0_J$ of C-AGB stars appears to be essentially constant within a certain ($J-K$)$_0$ range, so that C-AGB stars can be used as a standard candle. This so-called J-region AGB (JAGB) distance estimator \citep{Madore2020} is independent of, and complementary to, both the TRGB and Cepheid methods, and so provides a cross-check on them towards estimates of the local $H_0$. AGB evolution limits TP-AGB C stars to about $1.5 - 5 \Msun$, because at lower masses the third dredge up is never reached, while at higher masses, hot bottom burning fuses C to N \citep{Karakas2014, Marigo2017}.  

We emphasize study of the C star luminosity functions (LFs) in the near-infrared because the effects of extinction are much weaker, and because most recent C star or red giant LF studies have been performed using $JHK$-band photometry (e.g., \citealt{Marigo2003,Ripoche2020}) from the Two Micron All Sky Survey (2MASS; \citealt{2MASS}). Here, we present uniform and reliable samples of C giants in the LMC, SMC, Sgr and the MW in 2MASS CMDs and LFs.  We use extinction coefficients in the 2MASS bands from \cite{Gordon2003}, Table 4 and equations 4 - 7 of \citet{Ripoche2020} to derive absolute near-infrared magnitudes in the LMC and SMC.
To deredden 2MASS magnitudes in the Milky Way and for Sgr, we use the extinction coefficients from \citet{Yuan2013}.

\begin{figure*}
\epsscale{1.2}
\plotone{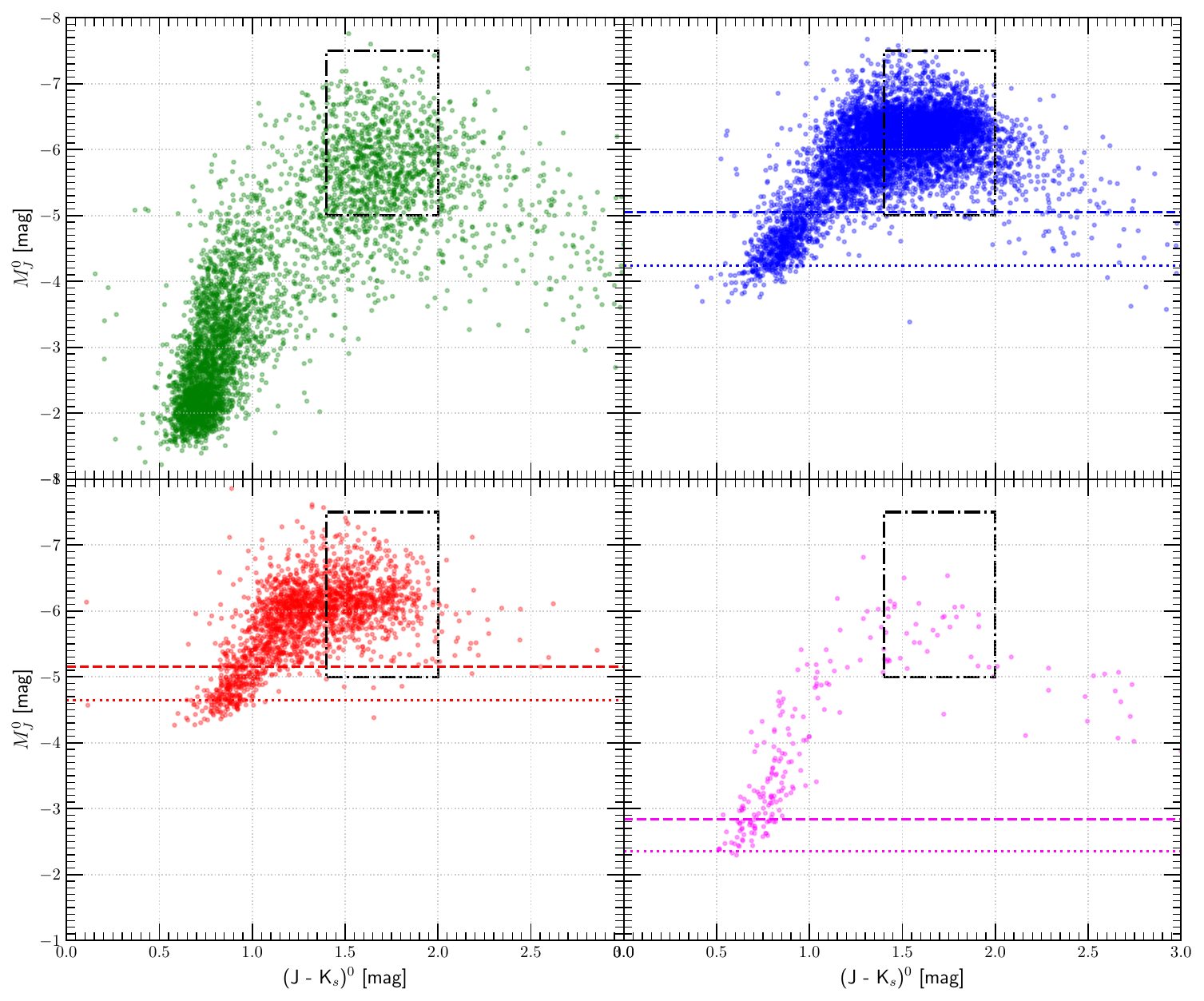}
\caption{Absolute dereddened $M_J^0$ magnitude vs ($J-K_s$)$^0$ color for C giants in the Milky Way (upper left), LMC (upper right), SMC (lower left) and Sgr (lower right). Single distance moduli are assumed for all stars in the LMC, SMC and Sgr (18.477, 18.977 and 17.0 respectively).  Our optical $G$-band apparent magnitude limit also limits the faintest absolute $M_J^0$ magnitudes in our samples. Limits are shown for $M_J^0$ based on our $G=16.5$ survey limit and the median and 95th percentile ($G-J$)$^0$ distributions for each of the dwarf galaxies (dashed and dotted lines, respectively).  The number of C stars shown for the Milky Way, LMC, SMC and Sgr is 2063, 6738, 2114 and 219, respectively.  For fitting the near IR CSLF, similar to the JAGB method of \citet{Ripoche2020}, we select C-AGB stars using both a color and an absolute magnitude cut:
$1.4<(J-K_s)^0<2$ and $-5 > M^0_J > -7.5$.  These boundaries are indicated by the dot-dashed black box in each plot.  The number of C-AGB stars within the box for the Milky Way, LMC, SMC and Sgr is  799, 3869, 763 and 35, respectively. }
\label{fig:JmKcmds}
\end{figure*}

Figure\,\ref{fig:JmKcmds} shows the de-reddened 2MASS CMDs for the C giants in the MW, LMC, SMC and Sgr.  A region of constant $M^0_J$, albeit with significant dispersion, is indeed evident within a limited ($J-K_s$) color range.  Within our survey sensitivity, the TPRGB is most obvious in the MW CMD, but separation is evident between the RGB and AGB in all four systems plotted. Similar to the JAGB method of \citet{Ripoche2020}, we select C-AGB stars using both a color and an absolute magnitude cut: $1.4<(J-K_s)^0<2$ and $-5 > M^0_J > -7.5$, marked in the figure by a dot-dashed box region. We fit the $J$-band LF for each system using both a Gaussian and Lorentz model, but the best-fit parameters are very similar either way.  Using the Bayesian Information Criterion \citep{Schwarz1978}, the Lorentzian is a superior fit for the LMC and SMC, as it better models the tails of the distribution. A Gaussian is only marginally a better fit for the MW, with similar results to the Lorentz fit.  Neither model is very good for Sgr, where by these criteria, there only 35 C-AGB stars.  We fit using 30 bins across the magnitude range, except for Sgr, where we use just 15 bins. 


Table\,\ref{tab:absJ_lorentz_fits} shows the model parameters for these $M^0_J$-band LFs. 
The mean  values for the LMC and SMC ($-6.30$ and $-6.15$, respectively) are very similar to those found by \citet{Ripoche2020} using only photometric identification of the C-AGB stars.  They attribute the fainter peak of the SMC to its lower metallicity, because dredge-up is more efficient at producing C$>$O when there is less O to begin with \citep{Iben1983}. Though there is broad internal dispersion, an age-metallicity relation, and metallicity gradients across both galaxies, the metallicity distributions differ, with the LMC having higher metallicity than the SMC:  [Fe/H]$_{LMC}\sim -0.4$ (e.g., \citealt{Smecker-Hane2002, Choudhury2016}), while [Fe/H]$_{SMC}\sim -0.7$ (e.g., \citealt{Povick2025, Mucciarelli2014}).  


The near-IR CSLF of the MW is considerably fainter and broader than for the LMC or SMC.  The fainter peak is consistent with higher metallicity of the progenitor population in the MW, predominantly thin and thick disk \citep{Abia2002,Abia2022}.  For Sgr, the uncertainty on the mean $\overline{M^0_J}$ is too large to make a meaningful comparison.  

\begin{deluxetable}{llrrcccc}
\tablecaption{Best fit parameters to $M_J^0$ luminosity functions}
\label{tab:absJ_lorentz_fits}
\tablehead{
\colhead{System} &
\colhead{Model} &
\colhead{No. of Stars} &
\colhead{Amplitude} &
\colhead{Mean} &
\colhead{Std.\ Dev.} &
\colhead{$\chi^2_{red}$} &
\colhead{BIC}
}
\startdata
MW  & Gaussian &  799 &  45.5 & $-5.69\pm 0.04$ & $0.67\pm 0.04$ &   1.155 &   41.39 \\
\ldots\  & Lorentz  &  \ldots\ &  48.0 & $-5.68\pm 0.04$ & $0.70\pm 0.06$ &   1.635 &   54.34 \\
\hline
LMC & Gaussian & 3869 & 403.3 & $-6.29\pm 0.01$ & $0.30\pm 0.01$ & 134.439 & 3640.07 \\
\ldots\ & Lorentz  & \ldots\ & 449.4 & $-6.31\pm 0.01$ & $0.27\pm 0.01$ &   8.057 &  227.74 \\
\hline
SMC & Gaussian &  763 &  68.3 & $-6.15\pm 0.01$ & $0.31\pm 0.01$ &  14.374 &  398.29 \\
\ldots\ & Lorentz  &  \ldots\ &  76.5 & $-6.15\pm 0.01$ & $0.29\pm 0.01$ &   0.829 &   32.57 \\
\hline
Sgr & Gaussian &   35 &   3.1 & $-5.61\pm 0.23$ & $0.59\pm 0.31$ &   1.705 &   28.58 \\
\ldots\ & Lorentz  &   \ldots\ &   3.0 & $-5.62\pm 0.26$ & $0.76\pm 0.56$ &   1.718 &   28.74 \\
\enddata
\end{deluxetable}

\begin{figure*}
\epsscale{1.1}
\gridline{\fig{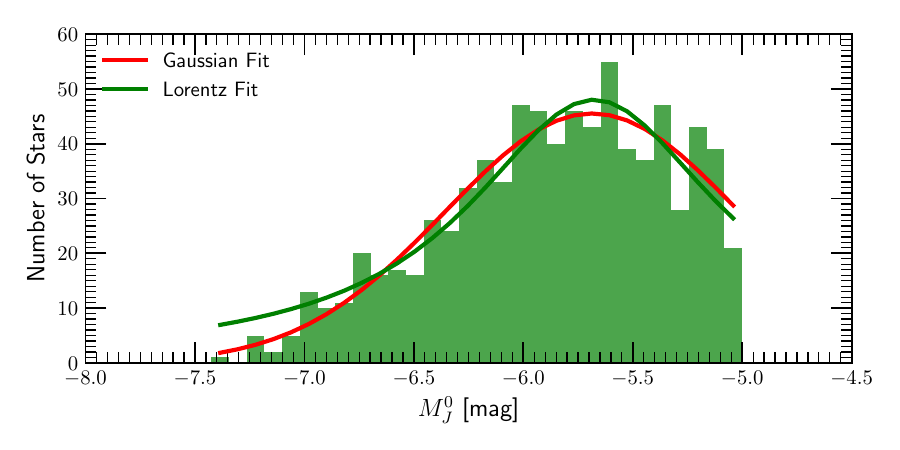}{0.5\textwidth}{(a) Milky Way} \fig{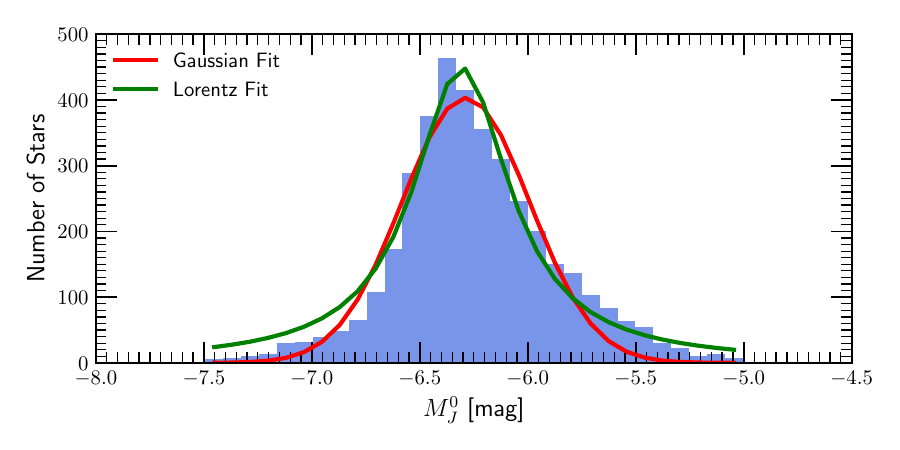}{0.5\textwidth}{(b) LMC}}
\gridline{\fig{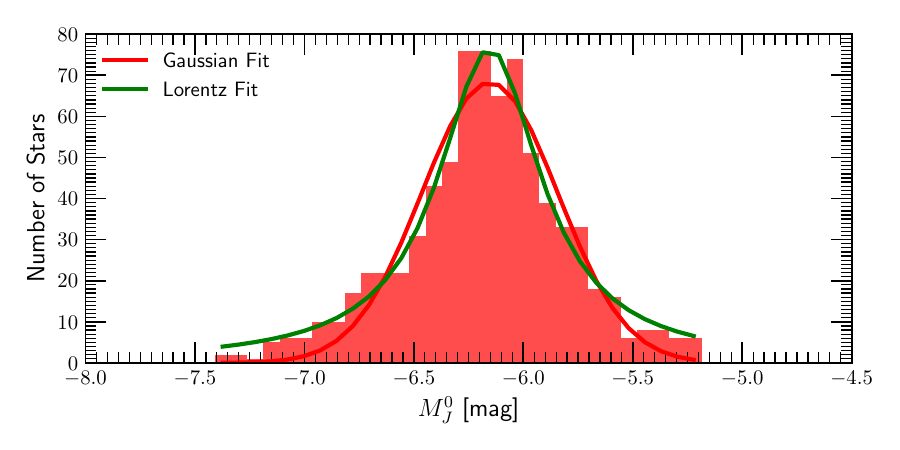}{0.5\textwidth}{(c) SMC}
  \fig{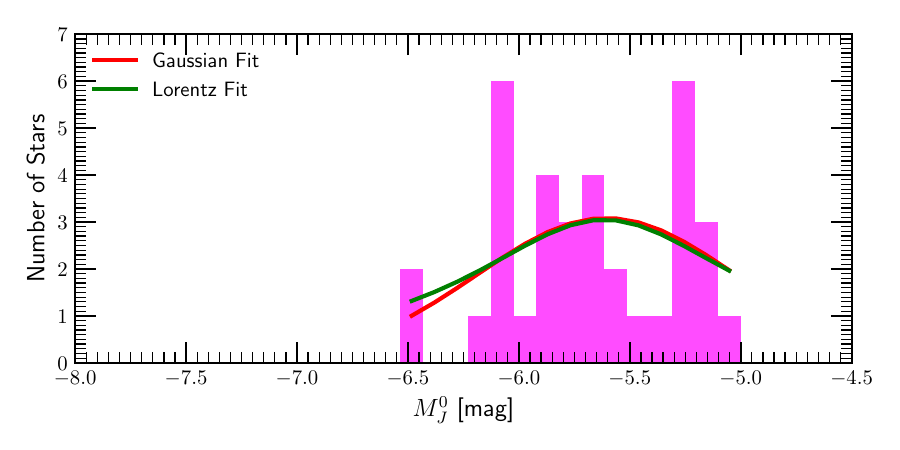}{0.5\textwidth}{(d) Sgr}}
\caption{Carbon star luminosity function, and model fits. Histograms of absolute $J$-band magnitudes are shown for C giants in the JAGB portions of the CMD highlighted in Figure\,\ref{fig:JmKcmds} for the MW (top left), LMC (top right), SMC (lower left) and Sgr (lower right).  Model fits to these distributions are shown as red line for Gaussian and green line for Lorentz models.  The best-fit parameters are listed in Table\,\ref{tab:absJ_lorentz_fits}.}
\label{fig:cslfs}
\end{figure*}

\section{Space Distribution of Milky Way C Giants \label{sec:spacedist}}

\citet{Roulston2025} studied the space distribution of dC stars, measuring their local space density to be $n_0 = 1.96^{+0.14}_{-0.12}\times10^{-6}\,\mathrm{pc}^{-3}$ (approximately one dC per local disk volume of radius 50\,pc), with a disk scale height of $H_z = 856^{+49}_{-43}\,$pc.  Given the low luminosities of main-sequence dCs ($5 < M_G < 9.5$), the $G \leq 16.5$ {\em Gaia} sample becomes incomplete for some dC stars beyond $\sim$250\,pc and for nearly all dC stars beyond $\sim$1500\,pc.  In contrast, the C giants studied here ($-5 < M_G < 0$) are all detectable in our sample out to roughly 20\,kpc, with the most luminous C-AGB stars accessible at nearly 200\,kpc.  This volume reach allows us to constrain multi-component spatial models that would be inaccessible to a dC-star sample.  The statistics of our C giant sample, however, support at most a two-component decomposition.

We study the distribution of all C giants combined, and we also fit the C-AGB and C-RGB subsamples separately.  We reason that the origin of excess carbon in C-AGB stars is intrinsic, arising from the upwelling of carbon-rich material during thermal pulsations on the AGB.  Non-AGB C giants, by contrast, are likely extrinsic: like the dC stars, they achieved atmospheric C$>$O through mass transfer from a former C-AGB companion.  Given the comparatively short lifetime of the red-giant phase ($t \approx 100$\,Myr for stars in the dC mass range; e.g., \citealt{Girardi2000}), carbon enrichment most probably occurred while the present-day red giant was still on the main sequence, so C-RGB stars are expected to be evolved dC stars analogous to the CH or CEMP-$s$ stars, with correspondingly high binary fractions.  A minority of C-RGB stars may be R-type stars, which are predominantly single and whose carbon enrichment is not yet fully understood, though stellar mergers have been proposed as an explanation \citep{Izzard2007, Zamora2009}.

We acknowledge that the complex formation and evolutionary history of the Milky Way halo defies characterization by the simple parameterizations we adopt.  A significant fraction of the halo is composed of stars accreted from disrupted satellite galaxies over cosmic time \citep[e.g.,][]{Johnston1996,Sharpe2024}, and unvirialized substructure such as streams and overdensities persists to the present day \citep{Belokurov2006}.  Even with our unprecedented sample of C giants, the statistics do not permit a reliable decomposition into more than two spatial components.

\begin{figure*}
\gridline{\fig{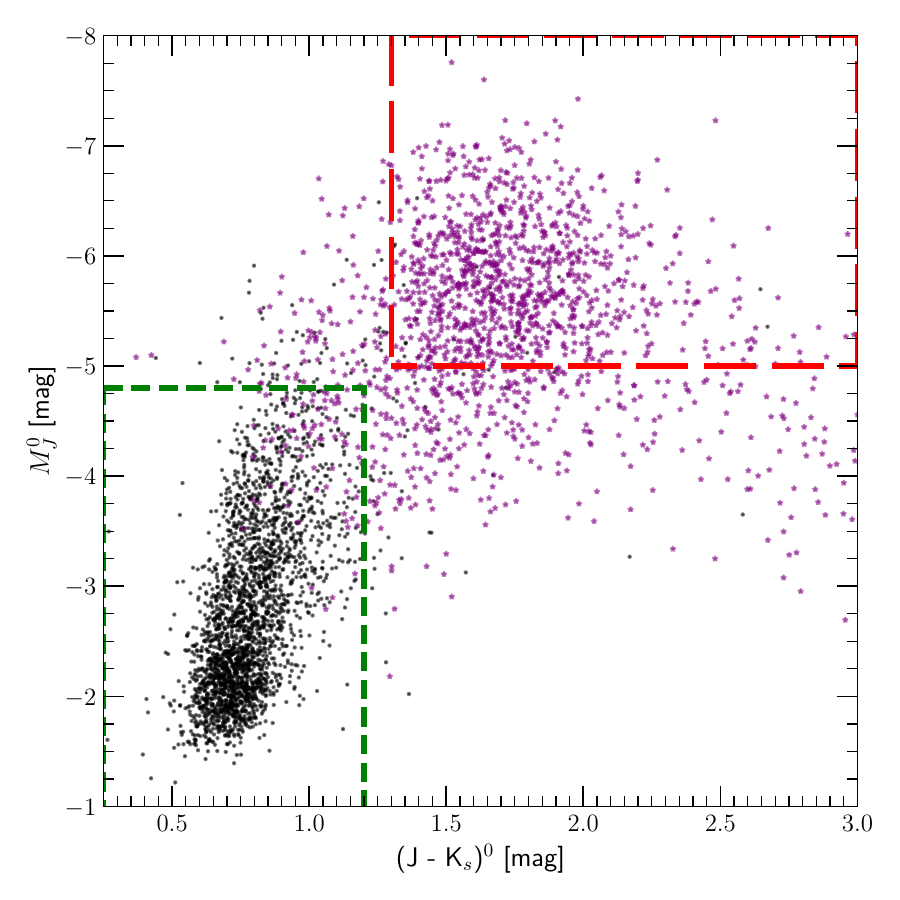}{0.5\textwidth}{(a) near-IR CMD}}
\gridline{
  \fig{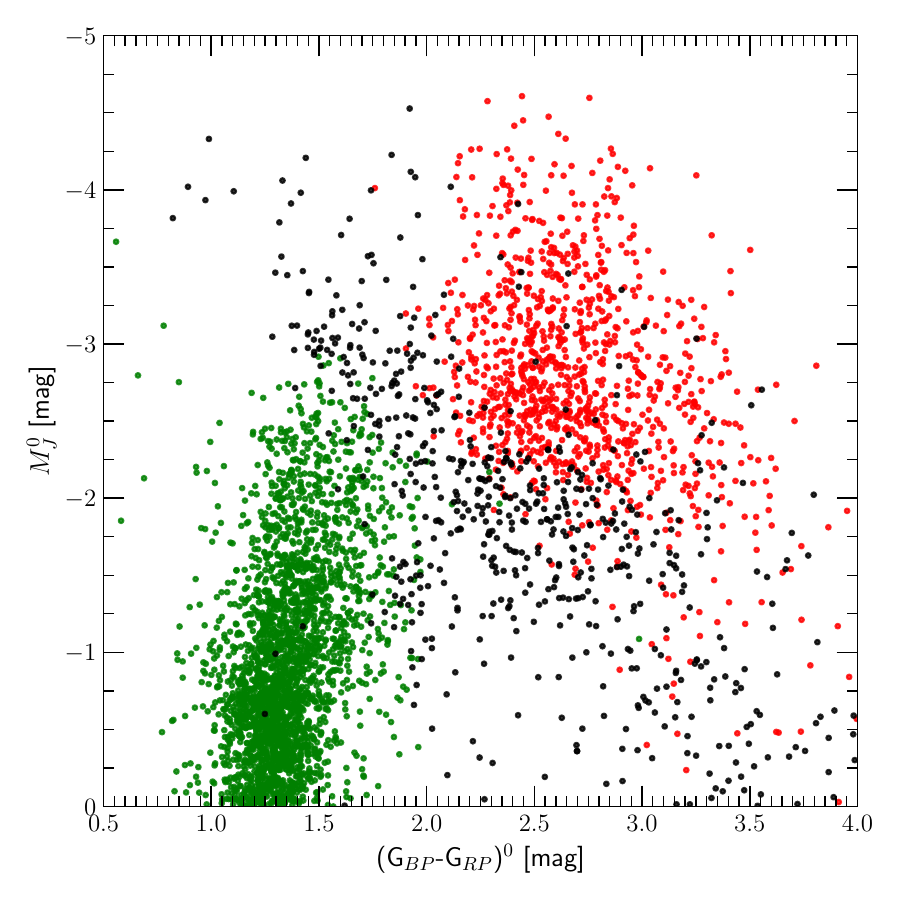}{0.5\textwidth}{(b) \textit{Gaia} CMD}
  \fig{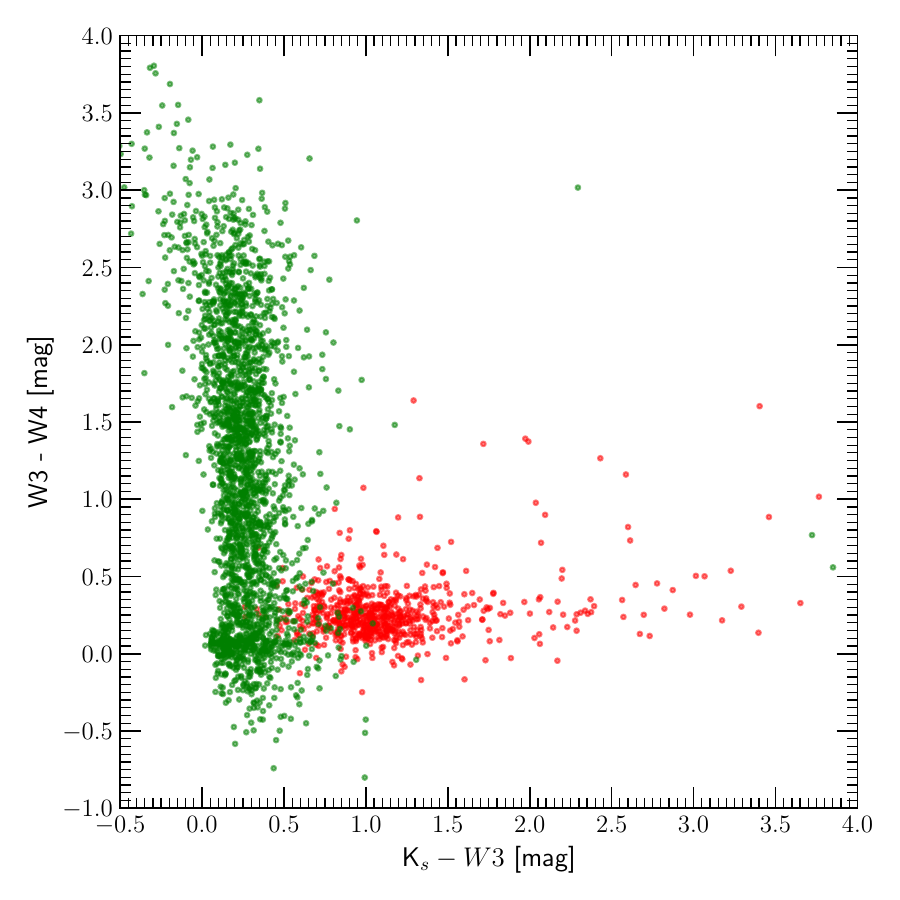}{0.5\textwidth}{(c) IR color-color diagram}}
\caption{Selection of clean C-AGB and C-RGB star samples from the Milky Way sample of C giants. 
 (a) In the near-IR CMD, we select non-contiguous regions for purity. For C-AGB stars, we require $M^0_J<-5$, and $(J-K_s)^0>1.3$ (red long-dash boundary). For C-RGB stars, we require $M^0_J>-4.8$, and $(J-K_s)^0<1.2$ (green short-dash boundary).  C-giants outside these regions have less clear classifications, as they are more likely to be affected by strong reddening and/or variability. Purplem stars points highlight the C giants matched to the \GDR3\, LPV catalog of \citet{Lebzelter2023}.  (b) The resulting samples from the selection in (a) are shown in the {\em Gaia} CMD.  C-AGB stars are shown in red, C-RGB in green, and remaining stars in black. (c)  Infrared color-color diagram for 3653 of our C giants matched to good quality ALLWISE photometry. Stars we classify as C-AGB (red points) are fairly well-separated, and extend redward in $K-W3$ along a track well-modeled by amorphous carbon dust shells of increasing optical depth \citep{Suh2024}.}
\label{fig:RGB_AGB}
\end{figure*}

In order to study the C-AGB and C-RGB stars separately, we examine the optical and near-IR (2MASS) CMDs for our Milky Way sample of 4276 C giants.  The near-IR CMD yields the cleanest separation with the least ambiguity.  For C-AGB stars, we require  $M^0_J<-5$, and $(J-K_s)^0>1.3$, yielding 1048 stars. For C-RGB stars, we require $M^0_J>-4.8$, and ($J-K_s$)$<1.2$, yielding 2611 stars. The 617 C giants outside these regions have less clear classifications, as they are more likely to be affected by strong reddening and/or variability. Figure\,\ref{fig:RGB_AGB}(a) shows these cuts in the near-IR CMD. 

Classical carbon stars are TP-AGB stars and are expected to be long-period variables (Miras, semiregulars, or irregular variables), with pulsation being an intrinsic characteristic of this evolutionary phase. We checked the frequency of long period variables (LPVs) against our C-AGB and C-RGB criteria by matching our Milky Way C giant sample to the \citet{Lebzelter2023} catalog of about a million $G<20$ LPVs selected from \GDR3.  From among our 4276 MW C giants, we find 1623 matches to this LPV catalog. In  Figure\,\ref{fig:RGB_AGB}(a), points show the LPVs as red points. Of our C-AGB stars, 98.3\% are LPVs, compared to 2.9\% of our C-RGBs.  This powerfully confirms the high purity of both our C-AGB and C-RGB samples, and also illustrates that C-AGB stars are essentially always LPVs. 

We also show in Figure\,\ref{fig:RGB_AGB}(b) the resulting samples in the optical {\em Gaia} CMD, where the separation is less clear, and even more dependent on reddening and variability.  

Mid-infrared photometry has often been used to study C stars, and long period variables (LPVs) in particular.  
We matched our Milky Way C giant sample to the {\tt gaiadr3.allwise\_best\_neighbour} table.
Of the 4277 Milky Way C giants in our sample, 3653 had good quality photometry from ALLWISE \citep{WISE, NEOWISE}.  In Figure\,\ref{fig:RGB_AGB}(c), we plot these stars in the $W3-W4$ vs. $K-W3$ color space.  Our C-RGB and C-AGB samples (green and red points, respectively) are relatively well separated.
The C-AGB sample extends redward in $K-W3$ along a track that is well-modeled by amorphous carbon dust shells of increasing optical depth \citep{Suh2024}.



\begin{deluxetable*}{llccc}
\tablecaption{Galactic Spatial Density Model Fits for Giant C Star
  Samples\label{tab:density_fits}}
\tablehead{
  \colhead{Model} &
  \colhead{Parameter} &
  \colhead{All C Stars} &
  \colhead{AGB C Stars} &
  \colhead{RGB C Stars} \\
  \colhead{} &
  \colhead{N$_*$} &
  \colhead{4,276} &
  \colhead{1048} &
  \colhead{2611}
}
\decimals
\startdata
\hline
\multicolumn{5}{l}{\textit{Sech}$^2$ \textit{disk}} \\
\hline
  & $n_{0,\mathrm{disk}}$\tablenotemark{a} ($10^{-8}$\,pc$^{-3}$)
    & $3.140^{+0.152}_{-0.146}$
    & $2.672^{+0.225}_{-0.210}$
    & $0.947^{+0.057}_{-0.055}$ \\
  & $H_z$ (pc)
    & $544.7^{+8.2}_{-8.1}$
    & $375.6^{+9.2}_{-9.1}$
    & $689.8^{+13.1}_{-12.8}$ \\
  & BIC
    & $-848.5$
    & $-1531.1$
    & $-1253.7$ \\
  & $\Delta$BIC
    & $782.9$
    & $43.8$
    & $473.4$ \\
\hline
\multicolumn{5}{l}{\textit{Exponential disk}} \\
\hline
  & $n_{0,\mathrm{disk}}$\tablenotemark{a} ($10^{-8}$\,pc$^{-3}$)
    & $7.388^{+0.352}_{-0.342}$
    & $6.897^{+0.496}_{-0.460}$
    & $2.143^{+0.131}_{-0.126}$ \\
  & $H_z$ (pc)
    & $310.0^{+4.6}_{-4.5}$
    & $206.3^{+4.0}_{-4.1}$
    & $397.4^{+7.7}_{-7.5}$ \\
  & BIC
    & $-1021.1$
    & $-1574.9$
    & $-1344.6$ \\
  & $\Delta$BIC
    & $610.3$
    & $0.0$
    & $382.5$ \\
\hline
\multicolumn{5}{l}{\textit{Double exponential disk}} \\
\hline
  & $n_{0,\mathrm{disk,1}}$\tablenotemark{a} ($10^{-8}$\,pc$^{-3}$)
    & $11.082^{+0.510}_{-0.490}$
    & $7.475^{+0.547}_{-0.514}$
    & $3.353^{+0.198}_{-0.189}$ \\
  & $H_{z,1}$ (pc)
    & $257.1^{+4.3}_{-4.2}$
    & $200.5^{+4.2}_{-4.3}$
    & $315.4^{+7.1}_{-7.1}$ \\
  & $n_{0,\mathrm{disk,2}}$\tablenotemark{a} ($10^{-8}$\,pc$^{-3}$)
    & $0.065^{+0.008}_{-0.007}$
    & $0.000^{+0.000}_{-0.000}$\tablenotemark{b}
    & $0.036^{+0.005}_{-0.004}$ \\
  & $H_{z,2}$ (pc)
    & $1146.4^{+38.6}_{-37.0}$
    & $3698^{+2805}_{-2316}$\tablenotemark{b}
    & $1339.0^{+54.0}_{-51.0}$ \\
  & BIC
    & $-1631.4$
    & $-1574.7$
    & $-1717.1$ \\
  & $\Delta$BIC
    & $0.0$
    & $0.1$
    & $10.0$ \\
\hline
\multicolumn{5}{l}{\textit{Sech}$^2$ \textit{disk + power-law halo}} \\
\hline
  & $n_{0,\mathrm{disk}}$\tablenotemark{a} ($10^{-8}$\,pc$^{-3}$)
    & $4.351^{+0.189}_{-0.184}$
    & $2.882^{+0.226}_{-0.213}$
    & $1.416^{+0.078}_{-0.075}$ \\
  & $H_z$ (pc)
    & $473.7^{+6.8}_{-6.7}$
    & $365.8^{+8.5}_{-8.4}$
    & $570.8^{+10.9}_{-10.7}$ \\
  & $n_{0,\mathrm{halo}}$\tablenotemark{a} ($10^{-11}$\,pc$^{-3}$)
    & $17.295^{+1.218}_{-1.184}$
    & $0.022^{+0.012}_{-0.008}$\tablenotemark{c}
    & $11.276^{+0.929}_{-0.883}$ \\
  & $\alpha$
    & $18.49^{+0.67}_{-0.65}$
    & $10.27^{+6.69}_{-6.92}$\tablenotemark{c}
    & $15.31^{+0.64}_{-0.61}$ \\
  & BIC
    & $-1562.6$
    & $-1540.9$
    & $-1724.4$ \\
  & $\Delta$BIC
    & $68.8$
    & $34.0$
    & $2.8$ \\
\hline
\multicolumn{5}{l}{\textit{Exponential disk + power-law halo}} \\
\hline
  & $n_{0,\mathrm{disk}}$\tablenotemark{a} ($10^{-8}$\,pc$^{-3}$)
    & $10.417^{+0.458}_{-0.443}$
    & $7.319^{+0.582}_{-0.550}$
    & $3.239^{+0.183}_{-0.175}$ \\
  & $H_z$ (pc)
    & $267.4^{+4.0}_{-3.9}$
    & $202.0^{+4.8}_{-4.7}$
    & $326.6^{+6.6}_{-6.4}$ \\
  & $n_{0,\mathrm{halo}}$\tablenotemark{a} ($10^{-11}$\,pc$^{-3}$)
    & $14.927^{+1.159}_{-1.108}$
    & $0.011^{+0.011}_{-0.011}$\tablenotemark{c}
    & $9.717^{+0.886}_{-0.846}$ \\
  & $\alpha$
    & $17.50^{+0.66}_{-0.63}$
    & $10.12^{+6.75}_{-6.86}$\tablenotemark{c}
    & $14.54^{+0.63}_{-0.61}$ \\
  & BIC
    & $-1627.3$
    & $-1573.8$
    & $-1727.1$ \\
  & $\Delta$BIC
    & $4.1$
    & $1.1$
    & $0.0$ \\
\enddata
\tablenotetext{a}{%
  Midplane space density propagated from the MCMC posterior on
  $\ln n_{0}$ with asymmetric errors.}
\tablenotetext{b}{%
  The second disk component is unconstrained for the AGB sample;
  the MCMC posterior on $n_{0,\mathrm{disk,2}}$ and $H_{z,2}$
  does not converge to a meaningful solution.}
\tablenotetext{c}{%
  Halo parameters are poorly constrained for the AGB sample,
  reflecting the limited vertical extent of AGB C stars.}
\tablecomments{%
  All parameter values are MCMC posterior medians with 68\% credible
  intervals. $\Delta\mathrm{BIC} = \mathrm{BIC} - \mathrm{BIC}_\mathrm{best}$
  is measured relative to the best-fitting model within each sample.
  The double exponential disk model has no halo component.}
\end{deluxetable*}

\begin{figure*}
\gridline{\fig{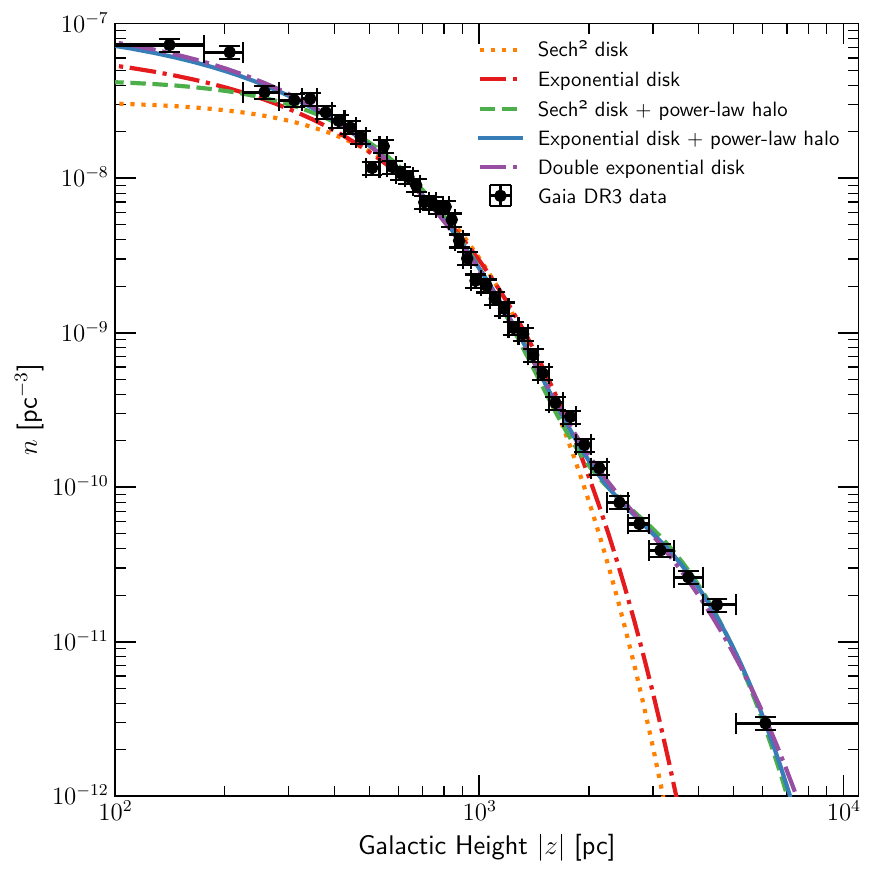}{0.5\textwidth}{(a) All C Stars}}
\gridline{
  \fig{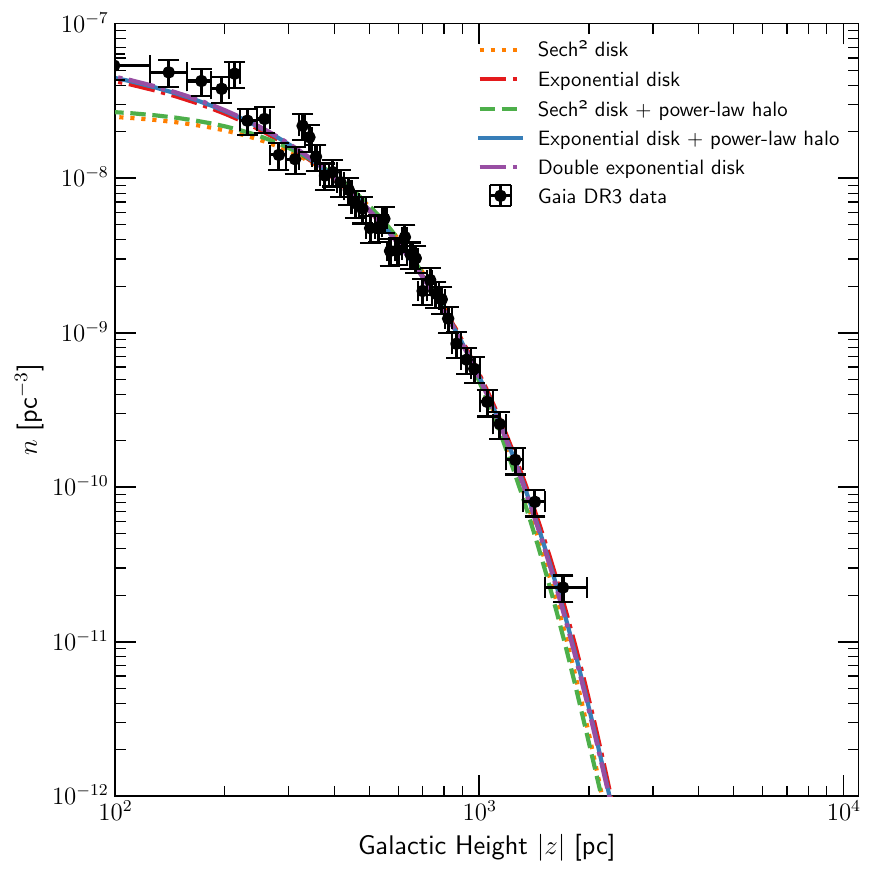}{0.5\textwidth}{(b) AGB C Stars}
  \fig{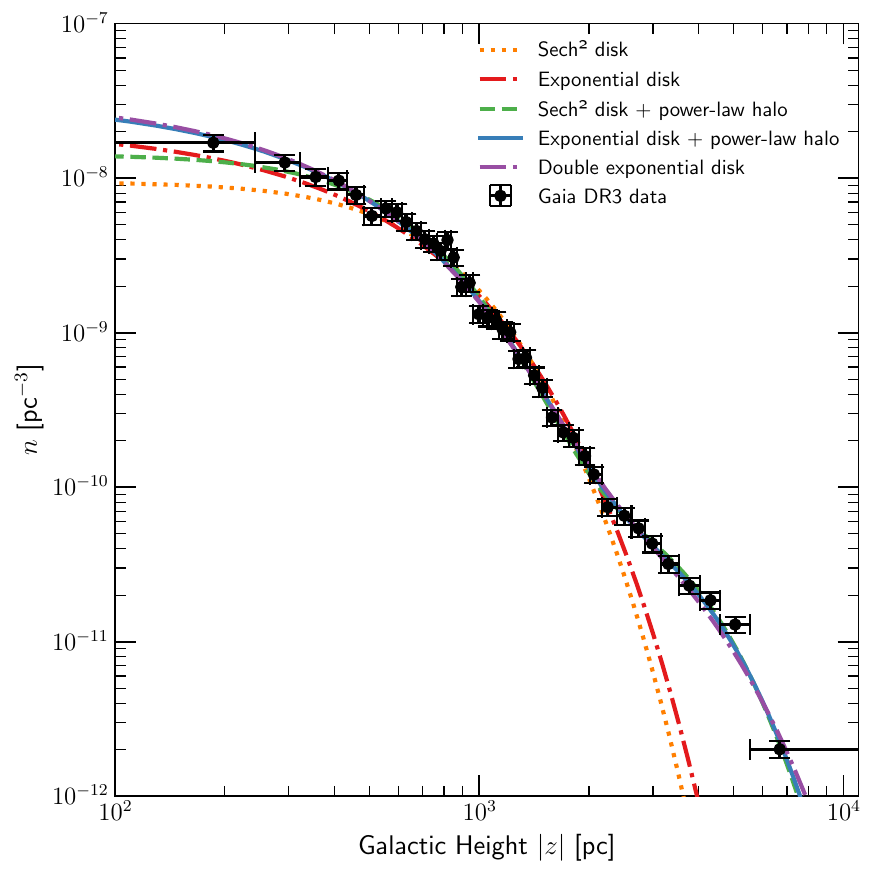}{0.5\textwidth}{(c) RGB C Stars}}
\caption{Stellar spatial number density as a function of Galactic height $|z|$ for (a) all C giants, (b) AGB C stars, and (c) RGB C stars in our \GDR3\, sample. Points show the 1/$V_\mathrm{max}$ density in each equal-$N$ $|z|$ bin with errors.  Five models fitted via MCMC are shown: a sech$^2$ disk (orange dotted), an exponential disk (red dot-dash), a sech$^2$ disk with a power-law halo (green dashed), an exponential disk with a power-law halo (blue solid), and a double exponential disk (purple dot-dash), with curves at the posterior median parameters. The all and RGB samples both are best fit with a two-population model, while the AGB sample is best fit with a single disk model.}
\label{fig:density_comparison}
\end{figure*}

\subsection{Model Selection \label{sec:modelsel}}

For each sample we fit five models to the binned 1/$V_\mathrm{max}$ density profile as a function of Galactic height $|z|$: a sech$^2$ disk, an exponential disk, each disk augmented with a power-law halo component, and a double exponential disk.  The disk profiles are

\begin{equation}\label{eq:sech2_model}
    n(z) = n_0\,\mathrm{sech}^2\!\left(\frac{|z|}{H_z}\right),
\end{equation}

\begin{equation}\label{eq:exp_model}
    n(z) = n_0\,e^{-|z|/H_z},
\end{equation}

\noindent where $n_0$ is the midplane density and $H_z$ is the scale height.  The halo component follows a power-law in the ellipsoidal galactocentric radius,

\begin{equation}
n_\mathrm{halo}(R,z) = n_{0,\mathrm{halo}}
    \left(\frac{r_q}{r_0}\right)^{\alpha},
\qquad
r_q = \sqrt{r^2 + \left(\frac{z}{q}\right)^{\!2}},
\end{equation}

\noindent where $r_q$ is the ellipsoidal radius, $q$ is a flattening parameter (we set $q=1$ for no flattening), $\alpha$ is the power-law slope, and $r_0 = R_\odot = 8.178$\,kpc is the solar galactocentric distance \citep{GRAVITYCollaboration2019}.  For the combined disk+halo models the total density is the sum of the disk and halo terms. The double disk models simply combine two models given in Equation \ref{eq:exp_model}, but with the requirement that one model have scale height $H_z<500$\,pc, and the other $H_z>500$\,pc.

All five models were fit by MCMC using \texttt{emcee} \citep{emcee} with 100 walkers and $110,000$ steps, discarding the first $10,000$ steps as burn-in. We adopt a chi-squared Gaussian likelihood and uniform priors on all parameters within broad physically motivated bounds.  Model selection uses the Bayesian Information Criterion (BIC), with $\Delta\mathrm{BIC} > 10$ considered strong evidence against the higher-BIC model.  Results for all samples and models are reported in Table~\ref{tab:density_fits}.

Each of the three samples was binned into 40 equal-$N$ vertical bins, giving average stellar counts of ${\sim}107$, ${\sim}26$, and ${\sim}65$ stars per bin for the All C Giants, AGB, and RGB samples respectively.  Volume corrections use the exact cone-plus-slab geometry of the $|b| > 5^\circ$ survey footprint following \citet{Roulston2025}, with the maximum detectable distance of each star computed from its dereddened absolute magnitude and the $G \leq 16.5$ survey limit.

\subsection{The Space Distribution of All Giant C Stars \label{sec:allspace}}

For the full C giant sample, disk-only models are decisively disfavored, with $\Delta\mathrm{BIC} > 600$ relative to either disk+halo model or the double exponential disk. The relative $\Delta\mathrm{BIC}$ between the two component models is smaller than that between disk only and two component. This suggests that while the data clearly show a disk only model is a poor fit, the selection between the two-component models is less conclusive. Based on the $\Delta\mathrm{BIC}$ values, we select the double exponential disk as our reported best fit, but we caution that this is not conclusive with $\Delta\mathrm{BIC}$ just 4.1 between double exponential disk and the exponential disk + power-law halo models. 


The preferred model is a double exponential disk, with a dominant thin-disk component of midplane density $n_{0,\mathrm{disk,1}} = 11.082\pm0.50\times10^{-8}\,\mathrm{pc}^{-3}$ and scale height $H_{z,1} = 257\pm4\,$pc, plus a much fainter, thicker second component with $n_{0,\mathrm{disk,2}} = 0.065\pm0.008\times10^{-8}\,\mathrm{pc}^{-3}$ and $H_{z,2} = 1146\pm38\,$pc. The combined midplane density, $n_{0,\mathrm{total}} = 11.147\pm0.51\times10^{-8}\,\mathrm{pc}^{-3}$, corresponds on average to approximately one C giant per spherical volume of radius 129~pc, or about 5.7\% the local space density of dC stars \citep{Roulston2025}.

\subsection{The Space Distribution of Carbon AGB Stars \label{sec:agbspace}}

For the C-AGB sample, all four models yield BIC values within a much narrower range than for the other samples. The simple exponential disk already returns the lowest BIC of any model considered, with the double exponential disk and the exponential disk + power-law halo offering only marginal improvements in likelihood ($\Delta\mathrm{BIC} = 0.1$ and $1.1$, respectively) despite each carrying two additional free parameters. Inspection of the posterior distributions disfavors these more complex models as well: in the double exponential disk, the second component is essentially unconstrained, while in the halo-bearing models both the halo normalization and power-law index are poorly constrained, with power-law terms of $\alpha \simeq 10\pm7$ and halo normalizations at least $10^5$ times lower than the disk component, consistent with zero. We therefore adopt the simple exponential disk as the preferred model for the C-AGB sample, both because it is favored by BIC and because it is the model with well constrained posterior distributions. 

The preferred exponential disk model yields a midplane density of $n_{0,\mathrm{disk}} = 6.9\pm0.5\times10^{-8}\,\mathrm{pc}^{-3}$ and a disk scale height of $H_z = 206\pm4\,$pc. This relatively small scale height is consistent with the intrinsic (self-enriched) origin of C-AGB stars; the thermal-pulse AGB phase is reached by intermediate-mass stars on timescales short enough that the population has not had time to kinematically heat to large vertical amplitudes. The midplane density corresponds on average to approximately one C-AGB star per spherical volume of radius $\sim151\,$pc, or about 3.5\% the local space density of dC stars \citep{Roulston2025}.

\begin{figure}
\centering
\includegraphics[width=\linewidth]{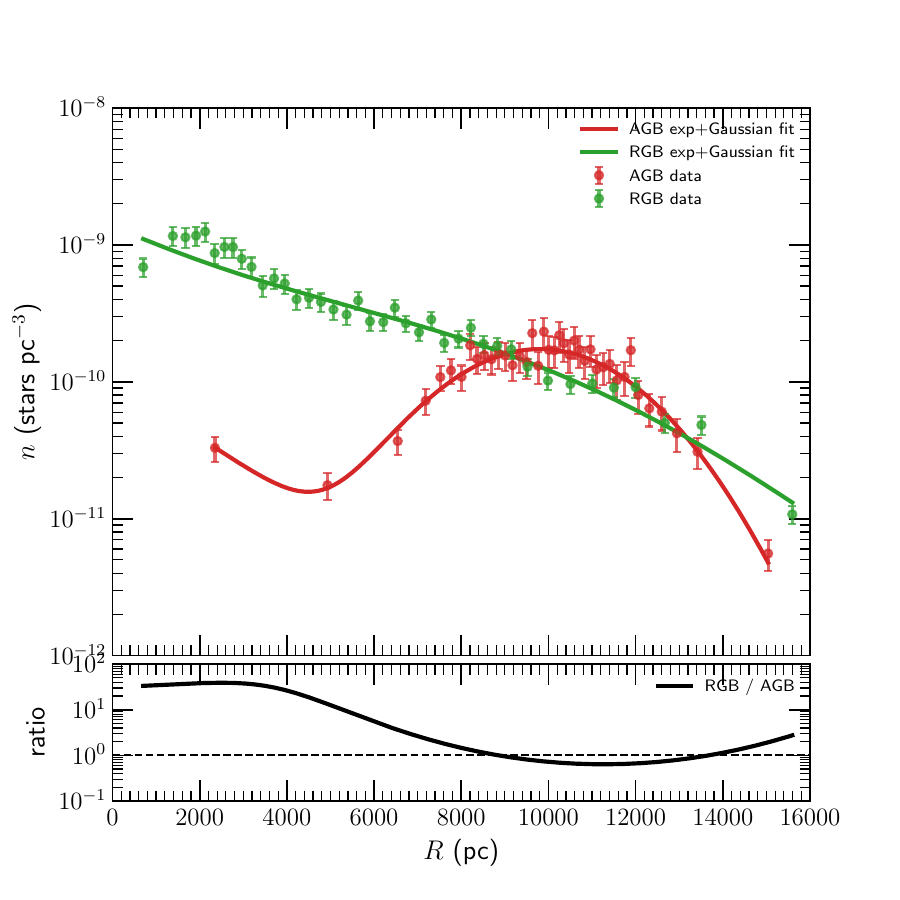}
\caption{Galactocentric radial density profile of Milky Way C-AGB (red) and C-RGB (green) stars, marginalized over $|z|$. TOP: $1/V_\mathrm{max}$ space density in 40 quantile-spaced $R$ bins, with best-fit exponential+Gaussian models (Equation \ref{eq:expgauss_model}) overplotted. The C-RGB profile declines smoothly, while the C-AGB profile shows a strong decrease for $R\lesssim 7$\,kpc and a peak at $R\sim 9.8$\,kpc, just outside the solar circle. BOTTOM: The C-RGB/C-AGB density ratio, which dips below unity near the C-AGB peak.}
\label{fig:Rprofile}
\end{figure}

The models above are all marginalized over Galactocentric radius $R $.  Motivated by the striking visual impression of changes in the AGB/RGB ratio with radius in Figure\,\ref{fig:XYRZ}, we now examine the radial distribution directly, which reveals an additional feature not captured by the vertical fits alone. Using the same $|b|>5^\circ$ C-AGB sample ($M^0_J<-5$, $(J-K_s)^0>1.3$; 1048 stars) and the analogous C-RGB sample (2611 stars; \S\ref{sec:spacedist}), we compute $1/V_\mathrm{max}$ space densities in 40 quantile-spaced $R$ bins, with uncertainties combining Poisson counting statistics and a Monte Carlo propagation of each star's parallax and photometric uncertainties, added in quadrature. Figure\,\ref{fig:Rprofile} shows the resulting radial profiles for both samples.

The C-RGB profile declines smoothly and nearly monotonically over the full radial range probed, consistent with the disk-dominated vertical fits of \S\ref{sec:rgbspace}. The C-AGB profile instead shows that C-AGB stars are rare between $R\sim2$--5\,kpc, followed by a rise to a well-defined peak near $R\sim 10$\,kpc, and a decline at larger $R$. We fit both profiles via weighted least-squares with a two-component model consisting of an exponential background plus a Gaussian excess,
\begin{equation}\label{eq:expgauss_model}
\rho(R) = A\,e^{-R/R_d} + B\,\exp\left[-\frac{(R-R_{\rm peak})^2}{2\sigma^2}\right].
\end{equation}
For the C-AGB profile this model is decisively preferred over simpler alternatives: relative to a plain exponential, $\Delta\mathrm{BIC}=427$, and relative to an unconstrained double exponential, $\Delta\mathrm{BIC}=289$, both well above the $\Delta\mathrm{BIC}>10$ threshold adopted in \S\ref{sec:modelsel}. We note that none of these models are selected on physical motivations, but rather on the expected continuous C-AGB decrease away from the Galactic center evident in  Figure\,\ref{fig:XYRZ}, with a possible density excess. The double-exponential fit is not a viable model; when both scale lengths are left completely free, the fit collapses to a degenerate solution in which the first term becomes an effectively flat background ($H_1\sim10^7$\,pc) and the second term's amplitude is consistent with zero, so that neither the dip nor the peak is reproduced.

The best-fit exponential+Gaussian model has background amplitude $A=(1.03\pm0.20)\times10^{-10}\,\mathrm{pc}^{-3}$ and scale length $R_d=2063\pm6\,$pc, with a Gaussian excess of amplitude $B=(1.73\pm0.09)\times10^{-10}\,\mathrm{pc}^{-3}$ ($B/A=1.68$) centered at $R_{\rm peak}=9820\pm85\,$pc and width $\sigma=1944\pm63\,$pc. At its peak, the Gaussian component exceeds the exponential background by a factor of $\sim$200, and $R_{\rm peak}$ lies $1642\,$pc outside the solar circle ($R_0=8178\,$pc; \citealt{GRAVITYCollaboration2019}).

We interpret this localized excess as a real feature of the C-AGB spatial distribution. Since C-AGB stars are intrinsically brighter than C-RGB stars, they are detectable to larger distances for the same apparent-magnitude cut.  The drop offs in C-AGB space density seen for $R\lesssim 7$\,kpc and $R\gtrsim 12$\,kpc are therefore not due to selection effects such as our magnitude limits.  If they were, then the less luminous C-RGBs would be affected even more strongly.  We note also that the C-AGB peak location is stable to sample subdivisions: it shifts by only $\sim165$\,pc when sightlines within $|l|<50^\circ$ of the Galactic center are excluded, disfavoring an origin in selection effects such as crowding or extinction, or physical environmental effects associated with the bulge/bar region.  Furthermore, the raw heliocentric distance distribution, prior to deprojection into Galactocentric $R$, does not show a corresponding peak-then-decline shape, disfavoring a simple detection-horizon or dust-map origin. 


The lower panel of Figure\,\ref{fig:Rprofile} shows the C-RGB-to-C-AGB density ratio, which falls from $\sim30$ near $R\sim2$\,kpc to a minimum below unity around $R\sim10$--11\,kpc, coincident with the C-AGB peak, before rising again to a few by the outer edge of our sample near $R\sim15$\,kpc. This crossover, in which the intrinsically rarer C-AGB stars locally match or exceed the C-RGB density, underscores how localized the enhancement is. Combined with the pronounced C-AGB deficit interior to $R\lesssim7$\,kpc discussed in \S\ref{sec:MW} -- there attributed to the inward increase in metallicity suppressing third dredge-up and to the greater age of the inner-disk population -- the radial profile suggests a more complete picture in which C-AGB stars are not simply depleted toward the Galactic center, but also modestly enhanced in an annulus just outside the solar circle. Plausible contributors include spatial structure in the star-formation history of the intermediate-age ($\sim0.5$--5\,Gyr; \S\ref{sec:cpops}) disk population from which C-AGB progenitors are drawn, or association with spiral-arm or other disk substructure in the outer Galaxy. Distinguishing among these possibilities will require kinematic information and comparison to the spatial distribution of other intermediate-age tracers, which we defer to future work.

\subsection{The Space Distribution of Carbon Red Giant Stars \label{sec:rgbspace}}

 For the C-RGB sample, disk-only models are again decisively disfavoured ($\Delta\mathrm{BIC} > 380$), and the exponential disk + power-law halo model is now the preferred model, with the sech$^2$ disk + power-law halo a close second at $\Delta\mathrm{BIC} = 2.8$. The detection of a halo component in the C-RGB sample, absent in the C-AGB sample, is consistent with the extrinsic (mass-transfer) origin of C-RGB stars; as evolved counterparts of dC stars, they share the old, kinematically hot stellar population that populates the Galactic halo.
 
 The preferred exponential disk + power-law halo model yields a midplane disk density of $n_{0,\mathrm{disk}} = 3.2\pm0.18\times10^{-8}\,\mathrm{pc}^{-3}$ and a scale height of $H_z = 327\pm6\,$pc. The halo has normalization $n_{0,\mathrm{halo}} = 9.7\pm0.9\times10^{-11}\,\mathrm{pc}^{-3}$ and power-law index $\alpha = 14.5\pm0.6$. We note that published studies typically find halo power-law density slopes ranging from $n \simeq 2.5$ to 5, but the inferred stellar-halo density profile depends significantly on the stellar tracer population, including age and metallicity selection (e.g., \citealt{Xue2015,Iorio2018,Mackereth2020}).

 The disk midplane density corresponds on average to approximately one C-RGB star per spherical volume of radius $\sim195\,$pc, roughly 1.7\% the local space density of dC stars \citep{Roulston2025}.


 The larger scale height of C-RGB stars ($H_z \approx 327$\,pc) compared to C-AGB stars ($H_z \approx 206$\,pc) supports the different formation pathways of these samples. C-AGB stars are intermediate-mass stars in a rapid evolutionary phase, while C-RGB stars are lower mass, old objects whose progenitor systems had time to acquire significant vertical velocities through disk heating. The C-RGB scale height also exceeds that of the general thin-disk population ($H_z \sim 200$--$300$\,pc; \citealt{Bovy2012, Mackereth2017}), suggesting that the mass-transfer binary channel that produces C-RGB stars preferentially draws from an older, more vertically extended stellar population, in the same way that dC stars were found to occupy a thick-disk-like distribution by \citet{Roulston2025}.

\section{Relationship Between C Star Populations  \label{sec:cpops}}

The measured midplane space densities of C-AGB, C-RGB, and dwarf carbon (dC) stars can be used to roughly constrain evolutionary connections among these populations, provided that differences in their characteristic lifetimes and vertical scale heights are taken into account. For a population $i$ in approximate steady state, the midplane density may be written as $n_i(0) \approx \Sigma_i/(2H_i) \propto \mathcal{R}_i \tau_i/H_i$, where $\mathcal{R}_i$ is the formation rate, $\tau_i$ is the observable lifetime, and $H_i$ is the scale height. Thus, the observed densities reflect the product of formation rate and lifetime rather than the formation rate alone.

Since dC stars originate from mass transfer in binary systems with C-AGB donors, the fraction of C-AGB stars that produce dCs may be expressed as
\begin{equation}
f_{\rm AGB \rightarrow dC}
\simeq
\frac{n_{\rm dC}}{n_{\rm C\mbox{-}AGB}}
\frac{\tau_{\rm C\mbox{-}AGB}}{\tau_{\rm dC}}
\frac{H_{\rm dC}}{H_{\rm C\mbox{-}AGB}}.
\end{equation}
The scale-height term arises because midplane space densities must be converted to surface densities, with $\Sigma \propto n(0)\,H$, in order to compare total populations and infer formation rates.  Given the large disparity between the short lifetime of the C-AGB phase ($\sim 10^6$ yr, depending on mass; \citealt{Kalirai2014, Marigo2017}) and the long lifetimes of dC stars (of order Gyrs), this fraction is expected to be small even if $n_{\rm dC} \gg n_{\rm C\mbox{-}AGB}$.

The local space density of dCs was measured by \citet{Roulston2025} to be 
$n_0\,\sim 2\times10^{-6}\,\text{pc}^{-3}$ (about one dC in every local disk volume of radius 50\,pc), with a relatively large disk scale height of $H_z\,\sim\,860\,$pc in the preferred $\text{sech}^2$ model (714\,pc in the less-favored exponential model).  Here, we measure the local space density of C-AGB stars to be $n_0\,\sim\,7\times10^{-8}\,\text{pc}^{-3}$, with a disk scale height of $H_z\,\sim\,206\,$pc.
If we reasonably assume lifetimes of $\sim\,10^6$yr for a C-AGB star and roughly of order $\sim 10$\,Gyr for a typical dC, we find that $\sim 1\%$ of C-AGB stars produce a dC.  
Given that about half of AGB stars are likely to be in binary systems (e.g., \citealt{Duchene2013} and references therein), this implies that about 2\% of C-AGB stars in binary systems will produce a dC via common-envelope evolution, Roche-lobe overflow, wind RLOF, or simple wind accretion.  While the dCs provide enduring evidence of their former C-AGB companions,  constraints on their accretion mechanisms are more elusive, requiring detailed modeling of the stellar masses, binary separations, mass outflow rates and the potentially rapid, unstable evolution of all these during mass transfer.


A related quantity is the fraction of dC stars that subsequently now
appear as carbon-rich red giant branch stars. This may be written as
\begin{equation}
f_{\rm dC \rightarrow C\mbox{-}RGB}
\simeq
\frac{n_{\rm C\mbox{-}RGB}}{n_{\rm dC}}
\frac{\tau_{\rm dC}}{\tau_{\rm C\mbox{-}RGB}}
\frac{H_{\rm C\mbox{-}RGB}}{H_{\rm dC}}.
\end{equation}
If we take a representative RGB lifetime of about 200\,Myr (e.g., \citealt{Girardi2000, Pietrinferni2004}), and using our space density fit results from \S\,\ref{sec:rgbspace}, we find about 30\% of dCs have evolved to C-RGB stars. 
Several competing factors affect this fraction and its interpretation.  Most dCs, especially lower mass examples, have not yet evolved up the red giant branch.  Carbon enhancements acquired through binary mass transfer may be diluted or modified during that evolution, lowering the RGB surface carbon abundance and the C/O ratio.  However, decreasing effective temperature on the RGB favors molecule formation, enhancing the detectability of C$_2$ and CN bands. Thus evolved dC descendants may remain spectroscopically identifiable as carbon stars if C/O remains above unity, while more strongly diluted systems may instead appear as CH, CEMP-$s$, or Ba-like stars with weaker or absent C$_2$ bands.  

The fraction of C-AGB stars that produce dwarf carbon (dC) stars, as defined above, accounts only for systems currently observed in the dC phase. However, a fraction (at least about 30\%) of these systems have already evolved off the main sequence and are now observed as carbon-rich red giant branch (C-RGB) stars. To account for this evolutionary flow, we include the descendant C-RGB population as part of the total yield of the dC formation channel. Using the same steady-state framework (i.e., space densities corrected by phase lifetimes and vertical scale heights), the contribution of this evolved population can be written as the product $f_{\rm AGB \rightarrow dC}\,f_{\rm dC \rightarrow C\mbox{-}RGB}$. The total fraction of C-AGB stars that have produced dC systems is therefore\begin{equation}f_{\rm AGB \rightarrow dC}^{\rm corr}=f_{\rm AGB \rightarrow dC}\left(1 + f_{\rm dC \rightarrow C\mbox{-}RGB}\right),\end{equation}which explicitly includes both systems presently observed as dCs and those that have subsequently evolved into the C-RGB phase; the result is about 1.3\%. Here, $f_{\rm dC \rightarrow C\mbox{-}RGB}$ should be interpreted as an effective descendant fraction inferred from the relative space densities, lifetimes, and scale heights of the two populations.

The range of initial stellar masses over which the third dredge-up efficiently produces carbon enrichment is fairly narrow, approximately $2 \lesssim M/M_\odot \lesssim 4.5$. At higher initial masses ($M \gtrsim 4.5\,M_\odot$), hot-bottom burning converts dredged-up carbon into nitrogen before it can be transported to the stellar surface. At lower masses ($M \lesssim 2\,M_\odot$), the stellar envelope is too small to sustain extended thermal-pulse evolution; the envelope is lost after only a few pulses, terminating evolution before the star can become carbon-rich. These effects together restrict the formation of carbon stars to a limited mass range, with a peak efficiency near $2 \lesssim M/M_\odot \lesssim 3$ \citep{Marigo2007}. This corresponds primarily to A-type stars, with typical lifetimes are 0.5 -- 2 Gyr.  C-AGB stars are therefore found predominantly in intermediate-age populations ($\sim 0.5$--$5$\,Gyr).  Additionally, C-AGB stars are found to have moderately metal-poor to solar metallicity, as it is more difficult to achieve C/O$>1$ at higher metallicities where the O content is high.

The space density of dCs measured by \citet{Roulston2025} used a sample selected from {\em Gaia} XP spectra in the same way as the C giant sample considered here.   Giants are all cool, so  C$_2$ and CN bands are easily detectable when C$>$O.  In dC systems, we know that the initial primary which became a C-AGB star had initial mass $2 \lesssim M/M_\odot \lesssim 3$, and we know that the dC had a lower mass than that, since it has not yet evolved off the main sequence.  The completeness of dC samples relative to all C$>$O dwarfs decreases for the warmest temperatures, since the C$_2$ and CN bands critical to selection become weaker; the warmest detected dCs have colors similar to G-type stars. The incompleteness may not be severe, since we know of the upper limit to the initial dC mass $< 2\,M_\odot $ imposed by the initial primary mass.  However, current dC masses are larger than their initial masses, since many dCs have since accreted a significant mass of C-rich material.

\section{X-ray Detected C Giants  \label{sec:xdets}}

Significant X-ray emission from giant stars is rare, so its detection can indicate interesting phenomena such as interacting binaries.  We therefore checked our catalog of high confidence C giants against catalogs from {\em Chandra}, {\em XMM-Newton}, and {\em eROSITA}. 

We first cross-matched our C giant catalog with the {\em Chandra} catalog (CSC v2.1,; \citealt{Evans2024}), since {\em Chandra} has the highest quality astrometry of any X-ray telescope.  

We note one {\em Chandra} detection 2CXO\,J051810.7-691604 associated with an LMC C star, which is a photometrically known LPV, OGLE\,LMC100.7 105748 \citep{Soszynski2009}.  The X-ray luminosity of the source is quite high - $L_X$(0.5-7\,keV)$=2.1\times10^{33}\lcgs$, far above the luminosity function of stellar coronal emission, and is most likely a wind-fed symbiotic star, with a WD accretor, similar to Draco C-1.

Another likely new LMC carbon symbiotic 2CXO\,J050717.4-684347, is detected at an X-ray flux corresponding to $L_X$(0.5-7\,keV)$=6.1\times10^{32}\lcgs$, and a known OGLE~LMC-LPV-19026 \citep{Soszynski2009}. However, the {\em Chandra} source catalog notes possible confusion in the X-ray detection.

We note that one C giant towards $\omega$~Cen (5.24\,kpc, \citealt{Soltis2021}) is a likely {\em Chandra} source.
Interestingly, this star is actually in the foreground of $\omega$~Cen, at about 3200\,pc \citep{Bailer-Jones2021}.  At $G=10.98$, this star (SOPS IV e-94, \citealt{Stock1972}) has a dereddened $M_G=-1.82$ and an X-ray flux of $f$(0.5-7\,keV)$=9.18\times 10^{-14}$\,\fcgs.  The X-ray source, 2CXO~J132601.5-473305, corresponds to a luminosity of $1.1\times 10^{32}$\lcgs.  Evolved, late-type giants are known to be very weak X-ray emitters (e.g., \citealt{Linsky1979, Ayres1981, Locatelli2025}), so this system is quite likely an accreting binary.  Such a C-RGB star likely once had a C-AGB companion, which is now the white dwarf accretor, so this symbiotic system should be variable. ASASSN-V\,J132601.59-473306.0 is indeed noted as an irregular variable in the VSX database.\footnote{We searched the International Variable Star Index (VSX) database, operated at AAVSO, Cambridge, Massachusetts, USA.}  

We matched the C giants in our LMC, SMC, MW and Sgr samples with the 4XMM\_DR14 catalog of X-ray detections \citep{Webb2020} from the {\em XMM-Newton} satellite mission \citep{Jansen2001}.  Using a 4$\arcsec$ search radius, we find 9 matches in the LMC and and 2 in the SMC with an 0.5-4.5\,keV $S/N>2$.  However, we find a similar number of matches after shifting the C giant positions by an arcminute, with a similar distribution of match separations and X-ray $S/N$.  We therefore conclude that crowding prevents a reliable identification of X-ray sources.  For Sgr, which is considerably less dense on the sky, there are no matches of C giants to X-ray sources in  4XMM\_DR14.

Among the MW C giants, only one XMM match is found - TW Hor.  At a distance of 460\,pc, the detected 0.5-4.5\,keV X-ray flux corresponds to log\,$L_X=30.11 \pm 0.04 \,\lcgs$.  TW Hor is a C-AGB star, a semi-regular variable with $M_G\sim -3.8$.  It is known to have UV excess and Fe UV line emission (\citealt{Ortiz2019} and references therein). 

A potential X-ray detection of TW Hor has been discussed before.  \citet{Schmitt2024} cross-matched red giants in \GDR3\, with the first all-sky X-ray scan (eRASS1) performed by the eROSITA X-ray instrument \citep{Predehl2021}.  Out of 1735 definitive red giants from \GDR3, they found 16 genuine red giant or supergiant X-ray sources. TW Hor was matched, but not considered valid because its bright magnitude ($G=4.54$) indicated likely optical contamination as the source of the detection.  However, we find that the {\em XMM-Newton} X-ray spectral energy distribution (SED) peaks at about 1\,keV, whereas the expected SED for optical contamination should be extremely soft (all counts below about 0.5\,keV).  We note that the X-ray luminosity of TW Hor is typical or a bit low compared to most X-ray detected AGB stars (c.f. \citealt{Ortiz2021}), indicating that it is unlikely to be a symbiotic system, as most of those have log\,$L_X \gtrsim 31$ (e.g., \citealt{Ball2025}).  The source of the X-ray emission may be coronal, but the mechanism for generating coronal emission in AGB stars is not clear, since plasma at the observed temperatures cannot be confined at such low gravities without a strong magnetic field, and strong fields are not observed in AGB stars. Strong magnetic fields are not expected in AGB stars, because they have deep convective envelopes and they rotate too slowly ($v\,sini< 10$\,km/s) to have significant dynamo activity (e.g., \citealt{Sahai2015}).  There are just a few dozen known AGB stars with detected X-ray emission \citep{Ortiz2021,Guerrero2024}.  A rapidly rotating, magnetically active main sequence companion could produce X-ray luminosities like this. Tidal locking in a tight orbit seems unlikely as the cause of rapid rotation because, given the large radius of the AGB star, the system would be in a common envelope mode.   Common envelope systems have eruptive, transient behavior and are short-lived (centuries at most; \citealt{Ivanova2013}), but TW Hor is a semi-regular variable with a period of about 158 days \citep{Samus2017}.   TW Hor could have an active companion spun up by accretion - either direct or wind Roche lobe overflow.  One simple check for binarity is the RUWE value for TW Hor, which is 1.43, indicating likely binarity.  Dwarf carbon (dC) stars are produced by accretion of C-rich material from a C-AGB star onto a main sequence companion.  Known dC stars are all consistent with being in binary systems with a white dwarf (the C-AGB remnant), and many dCs are known to have X-ray emission consistent with rapid rotation from either tidal locking or accretion-induced spin-up \citep{Roulston2019, Roulston2021, Roulston2022}.  TW Hor may be a system evolving towards a dC.  One known system farther along that path is the "Necklace", a planetary nebula with a hot central star in close orbit with a dC star \citep{Miszalski2013, Jones2026}.  




\citet{Salvato2025} matched eRASS1 to \GDR3\, using the NWAY algorithm \citep{Salvato2018}, which takes into account proximity on the sky, position errors, local source densities, and non-astrometric properties such as source fluxes, color, morphology, and sky motion.  They produce a matched catalog, which includes 0.2 -- 2.3\,keV fluxes, and match confidence flags from NWAY.  We choose only high confidence matches with  p\_any$>0.3$ and p\_i$>0.9$, and find 9 matches to Milky Way C giants.
However, these are bright stars, for which optical contamination may generate an X-ray detection.  After removing those contaminated objects with FLAG\_OPT=1 (which include TW Hor described above), we are left with two matches.  

The first such eRASS1 X-ray association (\GDR3\, 3229441606998725888, at $\alpha$$=$69.44015, $\delta$$=$$-1.31999$\,deg) is a luminous C-RGB star ($M_G= -2.42$, $M^0_J= -4.88$, $(J-K_s)^0=0.88$) at a distance of 10.2\,kpc \citep{Bailer-Jones2021}.  With an  $0.2-2.3$\,keV X-ray flux of $4.16\pm 0.58\,\fcgs$, its X-ray luminosity would be $5.2\pm 0.73 \times\,10^{33}$ erg s$^{-1}$.  This is in the expected range for symbiotic stars, so this object is worthy of further investigation, as, despite there being a few hundred known Galactic symbiotics, at most about a dozen contain carbon stars.  NEOWISE $W1$ and ZTF $r$-band light curves spanning 7 - 10 years shows variability of just a few tenths of a magnitude, and no strong outbursts.  The lack of strong variability rules out a C-AGB Mira donor, so this system is likely to be another interesting example of reversed accretion.  The current C-RGB star is extrinsic, i.e., it accreted C-rich material from a former AGB companion.  That companion is now the accretor in a symbiotic system.  Higher resolution optical spectra could confirm ionized gas expected in a symbiotic system via H$\alpha$, He\,II\,$\lambda 4686$, and Raman O\,VI\,$\lambda\lambda$6830,7088.  Detection of strong $s$-process element enhancements in e.g., Ba, Sr, Y, Zr, La, Ce, or Pb would support past AGB mass transfer, while technetium would indicate current, intrinsic dredge-up.

The second reliable eRASS1 X-ray association is a known C giant, TV Cen (\GDR3\, 6125811218911342848, at $\alpha$$=$183.63182, $\delta$$=$51.53263\,deg), a C-AGB star ($M^0_J= -6.15$, ($J-K$)$_0=1.68$) at a distance of 840\,pc \citep{Bailer-Jones2021}.  The ASAS-SN lightcurve shows strong $V$-band variability, and TV Cen is a semi-regular carbon variable (SR) with a pulsation period of about 264\,days \citep{Samus2017}.  With an  $0.2-2.3$\,keV X-ray flux of $3.04\pm 1.33\,\fcgs$, its X-ray luminosity is $2.5\pm 1.1 \times\,10^{30}$ erg s$^{-1}$.  Like TW Hor, this may be a system with a rapidly rotating main sequence companion, accreting C-rich material and evolving towards a dC. The {\em Gaia} RUWE value of 1.06 neither confirms nor refutes this binary hypothesis.  



\section{Summary  \label{sec:summary}}

Using the all-sky \textit{Gaia} DR3 catalog of high-confidence carbon stars of \citet{Roulston2025}, we have carried out the first homogeneous study of carbon giants throughout the Milky Way and nearby Local Group galaxies. Our principal conclusions are as follows.

For C giants in the Milky Way, we find that lower-luminosity carbon giants (C-RGB stars) exhibit a much larger fraction of elevated \textit{Gaia} DR3 RUWE values than C-AGB stars, providing strong evidence that most are extrinsic post-mass-transfer systems, likely evolved dwarf carbon stars.

Using the measured space densities of C-AGB, C-RGB, and dwarf carbon stars together with their characteristic lifetimes and scale heights, we estimate that only about 1\% of C-AGB stars produce a detectable dC star, rising to about 1.3\% once the descendant C-RGB population is included. This is broadly consistent with the $\sim 1\times10^{-6}\,\mathrm{pc}^{-3}$ dC space density predicted by early population synthesis models \citep{Kool1995} and with the comparably low binary mass-transfer efficiencies inferred for barium and CEMP-$s$ stars from population synthesis of similar mass-transfer channels \citep{Izzard2010, Abate2013}.

We identify 6937 carbon giants in the LMC, 2148 in the SMC, and 219 associated with the Sagittarius dwarf spheroidal, nearly quadrupling the previously known bright Sagittarius sample.

We show that the mean absolute $M^0_J$ magnitude of C-AGB stars in the LMC and SMC is nearly identical, confirming previous photometric studies and reinforcing their usefulness as distance indicators.

We show that C-AGB stars are strongly confined to the Galactic disk and exhibit a pronounced deficit inside $R \lesssim 7$~kpc, consistent with the combined effects of Galactic metallicity and age gradients.

Finally, several new candidate carbon symbiotic stars are identified through their X-ray counterparts, demonstrating that the \textit{Gaia} catalog provides an efficient means of discovering rare interacting binaries.

The \textit{Gaia} DR3 carbon star catalog transforms the study of Galactic carbon stars from investigations of heterogeneous compilations to analyses of a uniform, all-sky population. This enables the intrinsic and extrinsic carbon giant populations to be distinguished on a statistical basis and provides a foundation for future studies of binary evolution, TP-AGB evolution, Galactic structure, and carbon-star populations throughout the Local Group. In particular, detailed binary evolution modeling could address two open questions in turn: first, stellar evolution codes such as MESA \citep{Paxton2015} can constrain how much accreted mass, and what conditions of accretion rate and envelope mixing are required for a main-sequence star to become an observable dC star as it accretes C-rich material from a C-AGB companion; and second, binary population synthesis can map those thresholds onto the distribution of initial masses, mass ratios, and orbital separations capable of producing dCs and the C-RGBs that they later become.  Such modeling, combined with the uniform C giant samples presented here, would help identify which mass-transfer channel (e.g., common envelope, RLOF, WRLOF or simple wind accretion) dominates dC formation, clarifying the relationships between the intrinsic and extrinsic C star populations revealed by \textit{Gaia}.


\facility{Gaia, LAMOST, IRSA, WISE}

\software{Astropy \citep{astropy1, astropy2}, Matplotlib \citep{matplotlib}, Numpy \citep{numpy}, Scipy \citep{scipy}, Scikit-Learn \citep{Scikit-learn}, TOPCAT \citep{topcat}}

\begin{acknowledgments}
This study made use of data from the European Space Agency (ESA) mission {\it Gaia} (\url{https://www.cosmos.esa.int/gaia}), processed by the {\it Gaia} Data Processing and Analysis Consortium (DPAC, \url{https://www.cosmos.esa.int/web/gaia/dpac/consortium}). Funding for the DPAC
 has been provided by national institutions, in particular the institutions
 participating in the {\it Gaia} Multilateral Agreement.

This publication makes use of data products from the Wide-field Infrared Survey Explorer, which is a joint project of the University of California, Los Angeles, and the Jet Propulsion Laboratory/California Institute of Technology, and NEOWISE, which is a project of the Jet Propulsion Laboratory/California Institute of Technology. WISE and NEOWISE are funded by the National Aeronautics and Space Administration.

We are grateful for the Vizier service \citep{Vizier2000}, which we accessed for numerous purposes during this study.  This research made use of the SIMBAD database, operated at CDS, Strasbourg, France \citep{Simbad2000}

\end{acknowledgments}


\bibliography{main}{}
\bibliographystyle{aasjournal}



\end{document}

%% file: defs.tex
\def\GDR3{{\em Gaia }DR3}

\newcommand{\aox}{\ifmmode{\alpha_{\mathrm{ox}}} \else $\alpha_{\mathrm{ox}}$\fi} 
\newcommand{\atoms}{\ifmmode{\mathrm{\,atoms~cm^{-2}}} \else \,atoms cm$^{-2}$\fi}
\newcommand{\ax}{\ifmmode{\alpha_x} \else $\alpha_x$\fi} 
\newcommand{\bprp}{\ifmmode{G_{BP}-G_{RP}} \else $G_{BP} - G_{RP}$\fi} 
\newcommand{\cmsq}{\ifmmode{\mathrm{cm^{-2}}} \else cm$^{-2}$\fi}
\newcommand{\degsq}{\ifmmode {\mathrm{deg^2}} \else deg$^2$\fi}
\newcommand{\perdegsq}{\ifmmode {\mathrm{deg^{-2}}} \else deg$^{-2}$\fi}
\newcommand{\ew}{\ifmmode{W_{\lambda}} \else $W_{\lambda}$\fi}
\newcommand{\fbol}{\ifmmode f_{\mathrm{bol}} \else $f_{\mathrm{bol}}$\fi} 
\newcommand{\fcgs}{\ifmmode \mathrm{erg~cm^{-2}~s^{-1}}\else erg~cm$^{-2}$~s$^{-1}$\fi}
\newcommand{\lcgs}{\ifmmode \mathrm{erg~s^{-1}}\else erg~s$^{-1}$\fi}
\newcommand{\flamcgs}{\ifmmode \mathrm{erg\,cm^{-2}\,s^{-1}\,\AA^{-1}}\else erg\,cm$^{-2}$\,s$^{-1}$\,\AA$^{-1}$)\fi}
\newcommand{\fnucgs}{\ifmmode {\mathrm{erg~cm^{-2}~s^{-1}~Hz^{-1}}}\else erg~cm$^{-2}$~s$^{-1}$~Hz$^{-1}$\fi}
\newcommand\Ha{\ifmmode {\mathrm H}\alpha \else H$\alpha$\fi}
\newcommand\Hb{\ifmmode {\mathrm H}\beta \else H$\beta$\fi}
\newcommand{\kms}{\ifmmode~{\mathrm{km~s}}^{-1}\else ~km~s$^{-1}~$\fi}
\newcommand{\lnucgs}{\ifmmode erg~s^{-1}~Hz^{-1}\else erg~s$^{-1}$~Hz$^{-1}$\fi}

\newcommand{\logz}{\ifmmode{\mathrm{log}}~z \else log$~z$\fi}
\newcommand{\lo}{\ifmmode l_o \else $~l_o$\fi}
\newcommand{\Lo}{\ifmmode L_o \else $~L_o$\fi}
\newcommand{\lx}{\ifmmode l_x \else $~l_x$\fi}
\newcommand{\Lx}{\ifmmode L_x \else $~L_x$\fi}
\newcommand{\lbol}{\ifmmode L_{\mathrm{bol}} \else $L_{\mathrm{bol}}$\fi}
\newcommand{\Lbol}{\ifmmode L_{\mathrm{bol}} \else $L_{\mathrm{bol}}$\fi}
\newcommand{\LBol}{\ifmmode L_{\mathrm{bol}} \else $L_{\mathrm{bol}}$\fi}
\newcommand{\LEdd}{\ifmmode L_{\mathrm{Edd}} \else $L_{\mathrm{Edd}}$\fi}
\newcommand{\Lsun}{\ifmmode {L_{\odot}}\else${L_{\odot}}$\fi}
\newcommand{\LxLbol}{\ifmmode L_x/L_{\mathrm{bol}} \else $L_x/L_{\mathrm{bol}}$\fi}
\newcommand{\rEdd}{\ifmmode L/L_{\mathrm{Edd}} \else $L/L_{\mathrm{Edd}}$\fi}
\newcommand{\REdd}{\ifmmode L/L_{\mathrm{Edd}} \else $L/L_{\mathrm{Edd}}$\fi}
\newcommand{\Rblr}{\ifmmode {R_{\mathrm BLR}} \else $R_{\mathrm BLR}$\fi}
\newcommand{\lamEdd}{\ifmmode \lambda_{\mathrm{Edd}} \else $\lambda_{\mathrm{Edd}}$\fi}
\newcommand{\mbh}{\ifmmode {M_{\rm BH}}\else${M_{\rm BH}}$\fi}
\newcommand{\Mbh}{\ifmmode {M_{\rm BH}}\else${M_{\rm BH}}$\fi}
\newcommand{\mdot}{\ifmmode \dot{m} \else $\dot{m}$\fi}
\newcommand{\mdote}{\ifmmode \dot{m}_{E} \else $\dot{m}_{E}$\fi}
\newcommand{\mone}{\ifmmode ^{-1}\else$^{-1}$\fi}
\newcommand{\msun}{\ifmmode {M_{\odot}}\else${M_{\odot}}$\fi}
\newcommand{\Msun}{\ifmmode {M_{\odot}}\else${M_{\odot}}$\fi}
\newcommand{\mtwo}{\ifmmode ^{-2}\else$^{-2}$\fi}
\newcommand{\Mvir}{\ifmmode {M_{\rm BH}^{\mathrm SE}}\else${M_{\rm BH}^{\mathrm SE}}$\fi}
\newcommand{\nhgal}{\ifmmode{ N_{H}^{Gal}} \else N$_{H}^{Gal}$\fi}
\newcommand{\nh}{\ifmmode{\mathrm N_{H}} \else N$_{H}$\fi}
\newcommand{\nhintr}{\ifmmode{ N_{H}^{intr}} \else N$_{H}^{intr}$\fi}
\newcommand{\nhtot}{\ifmmode{ N_{H}^{tot}} \else N$_{H}^{tot}$\fi}
\newcommand{\nhz}{\ifmmode{ N_{H}^z} \else N$_{H}^z$\fi}
\newcommand{\oi}{\ifmmode{\mathrm [O\,II]} \else [O\,II]\fi}
\newcommand{\oii}{\ifmmode{\mathrm [O\,II]} \else [O\,II]\fi}
\newcommand{\oiii}{\ifmmode{\mathrm [O\,III]} \else [O\,III]\fi}
\newcommand{\optebl}{\ifmmode L_{\rm 2500\,\AA} \else $~L_{\rm 2500\,\AA}$\fi}
\newcommand{\opteml}{\ifmmode l_{\mathrm{2500\,\AA}} \else $~l_{\mathrm{2500\,\AA}}$\fi}
\newcommand{\rhodC}{\ifmmode{ \rho_{\mathrm{dC}}} \else $\rho_{\mathrm{dC}}$ \fi}
\newcommand{\Teff}{\ifmmode T_{\mathrm{Eff}} \else $T_{\mathrm{Eff}}$\fi}
\newcommand{\xebl}{\ifmmode L_X \else $~L_X$\fi}
\newcommand{\xeml}{\ifmmode l_{\mathrm{2\,keV}} \else $~l_{\mathrm{2\,keV}}$\fi}

\def\geqsim{\lower.73ex\hbox{$\sim$}\llap{\raise.4ex\hbox{$>$}}$\,$}
\def\leqsim{\lower.73ex\hbox{$\sim$}\llap{\raise.4ex\hbox{$<$}}$\,$}

\newcommand{\umg}{\ifmmode{\mathrm{(}u-g\mathrm{)}} \else ($u-g$)\fi}
\newcommand{\gmr}{\ifmmode{\mathrm{(}g-r\mathrm{)}} \else ($g-r$)\fi}
\newcommand{\rmi}{\ifmmode{\mathrm{(}r-i\mathrm{)}} \else ($r-i$)\fi}
\newcommand{\gmi}{\ifmmode{\mathrm{(}g-i\mathrm{)}} \else ($g-i$)\fi}
\newcommand{\imz}{\ifmmode{\mathrm{(}i-z\mathrm{)}} \else ($i-z$)\fi}
\newcommand{\jmh}{\ifmmode{\mathrm{(}J-H\mathrm{)}} \else ($J-H$)\fi}
\newcommand{\hmk}{\ifmmode{\mathrm{(}H-K\mathrm{)}} \else ($H-K$)\fi}
\newcommand{\ctwo}{\ifmmode C_2 \else C$_2$\fi}